\documentclass[a4paper,11pt]{article}
\usepackage{jcappub} 
\usepackage{lineno}

\newcommand{\arcsinh}{\text{arcsinh}}

\usepackage{rotating}
\usepackage{xcolor}

\arxivnumber{2508.18434} 
\title{\boldmath Rotating Neutron Stars with Dark Matter Halos}

\author[a]{Shafayat Shawqi}
\author[b]{Andreas Konstantinou}
\author[a]{Sharon M. Morsink}
\affiliation[a]{Department of Physics, University of Alberta, Edmonton, AB, T6G 2E1, Canada}
\affiliation[b]{Department of Physics, University of Cyprus, P.O. Box 20537, 1678 Nicosia, Cyprus}

\emailAdd{shawqi@ualberta.ca}
\emailAdd{akonst29@ucy.ac.cy}
\emailAdd{morsink@ualberta.ca}

\abstract{If dark matter (DM) exists in halos around rotating neutron stars (NSs), it will be essential to understand the effects of rotation on the distribution of DM and baryonic matter (BM) in the stars to interpret observations. In this work, we construct rapidly rotating dark matter admixed neutron stars (DANS) with DM halos using the two-fluid approximation, where the BM and DM interact only through gravity. Our goal is to describe rapidly rotating millisecond-period DANS spun up by the accretion of BM from a zero angular momentum state. We extend the Rapidly Rotating Neutron Star ({\tt\string RNS}) code to compute axisymmetric configurations in which the BM rotates rigidly while the DM remains torque-free and differentially rotates through the frame-dragging of spacetime.  For the first time, we examine in detail local and global definitions of mass in general relativity for two-fluid systems, showing how their differences affect the interpretation of baryonic and dark component masses. We compute energy density and frame-dragging frequency profiles for DANS with three different characteristic DM halos. We demonstrate that rapid BM rotation reduces DM halo sizes if central energy densities are kept constant between non-rotating and rotating models. We also construct sequences of DANS to create mass and radius curves and compare rotating and non-rotating cases. Finally, we quantify deviations in the spacetime metric outside the baryonic surfaces of these sequences of stars caused by the DM halos. We hypothesize that the size of this quantity could indicate whether a DM halo will significantly impact X-ray pulse profile modeling. These results provide a framework for assessing the observational consequences of DM halos around rapidly rotating NSs.}

\begin{document}
\maketitle
\flushbottom
\newcommand{\AK}[1]{\textcolor{magenta}{ AK: #1}}

\section{Introduction}\label{sec:intro} 

Neutron stars (NSs) in closely interacting binary systems can accrete material from their companion and spin up to relativistic speeds due to conservation of angular momentum. The fastest spinning NS observed, PSR J1748-2446ad, is attributed to have such a recycled spin-up process \citep{Hessels2006}. These NSs can release energy by emitting X-rays from their surfaces. Observations by the Neutron Star Interior Composition ExploreR (NICER) of pulsed X-ray emission from five such recycled NSs, PSR J0030+0451 \citep{Riley2019, Miller2019}, PSR J0740+6620 \citep{Salmi2024a, Dittmann2024}, PSR J0437-4715 \citep{Choudhury2024}, PSR J1231-1411 \citep{Salmi2024b}, and PSR J0614-3329 \citep{Mauviard2025}, along with comparisons with relativistic models for the pulse profiles \citep{Bogdanov2019b}
have been used to infer their masses and radii.

All past analyses of pulsed X-ray emission from rotating NSs have made the natural assumption that no dark matter (DM) exists in the NS. However, some DM models allow for the existence of dark matter admixed neutron stars (DANS) where a small fraction of the compact object's mass is composed of dark matter. Accumulation of DM in NSs can occur through accretion from the galactic DM halo \citep{Goldman1989, Kouvaris2008, Kouvaris2010, Nelson2019}, production of DM from Standard Model species due to extreme core densities \citep{Motta2018a, Ellis2018, Baym2018, McKeen2018, Motta2018b}, or high temperatures in young NSs \citep{Ellis2018, Nelson2019, Reddy2022}, NS mergers with hypothesized dark stars \citep{Kouvaris2015, Ciarcelluti2011}, or baryon accretion in dark stars \citep{Kamenetskaia2022}. 

Depending on the properties and abundance of the dark matter particles, the resulting DANS can have one of two types of configurations. The dark matter in a core DANS is concentrated inside the outer radius of the baryonic matter (BM). A halo DANS has dark matter distributed both inside and outside of the baryonic surface. Recent work \citep{Rutherford2023, Rutherford2024} has explored how to analyze NICER data if the observed pulsars are DANS with DM cores, but a better understanding of how rotation affects the properties of DANS is required to extend this type of analysis to DANS with DM halos. 

In this paper, we construct rapidly rotating DANS that have acquired their angular momentum through the accretion of baryonic matter from a companion star. We employ the two-fluid approximation, where the baryonic and dark matter are assumed to interact only through gravity. While the two-fluid approximation is not a necessary choice, it provides a simple method for determining the surface of the baryonic component of the star, which is where the observed X-rays are emitted from. We assume that the DANS is born with zero angular momentum. The accreted plasma exerts a torque on the baryonic component of the star, causing it to spin faster as in standard pulsar recycling \citep{1977ApJ...217..578Ghosh}. Since there is no assumed coupling between the baryonic and dark matter, the accreted material does not exert a torque on the DM component, meaning that the angular momentum of the DM in the DANS continues to vanish. However, the angular momentum of the DANS is non-zero due to the rotation of the baryonic material and acts as a source for the dragging of inertial frames. As a result, the DM will acquire a non-zero differential angular velocity even though the DM has zero angular momentum. 

There are a few earlier computations of rotating DANS that make a variety of different assumptions. One approach is to assume a type of interaction between the standard model and DM sectors and solve the field equations with the stress tensor for the full field theory in either the slow rotation approximation \citep{Routaray2025} or rapid rotation \citep{Guha2021}. Another approach is to use the two-fluid approximation, as was done in the slow-rotation approximation \citep{Cronin2023} and for rapid rotation \citep{Mukhopadhyay2017, Konstantinou2024, Cipriani2025a, Issifu2025, Cipriani2025b}. In all of these cases, the dark matter is assumed to rotate rigidly with an arbitrary angular velocity independently of the baryonic matter rotation rate. Additionally, computations of rotating fermion-boson stars \citep{2024PhRvD.110l3019Mourelle} have some similarities to the two-fluid computations, where the bosonic scalar field component plays the role of a DM field that extends to infinity.

In general, whether DM in a rotating DANS possesses non-zero angular momentum or not is an open question. The answer depends on the DM accumulation process in the NS. If DM is accreted by the NS from the galactic DM halo, asymmetric accretion \citep{Spergel1988} may add non-zero angular momentum to the DM in the DANS. If DM particles are produced from Standard Model species in a rotating star, angular momentum from the preceding structure may be transferred to the DM in the resulting two-fluid configuration. Another possible scenario is the merger of two DANSs in a binary, where the DM component in the rotating remnant retains angular momentum from the binary \citep{Giangrandi2025}.

In this paper, however, our only focus is the scenario of accretion-powered pulsars. These are old pulsars in which the BM has had enough time to spin down and lose most of its angular momentum, before being spun-up by accretion of BM from their binary companions. In such a scenario, the main process through which the DM fluid could possess non-zero angular momentum is through angular momentum retention from the progenitor of the NS. Through Monte-Carlo population synthesis, ref.~\citep{Giguere2006} found that birth spin distribution of pulsars can be described by a Gaussian with a mean of 300 ms and a standard deviation of 150 ms. Such rotation rates are within the realm of very slow rotation. For example, even in the case of a 100 ms spin period, the dimensionless angular momentum parameter $J/M^2$ (ie. the dimensionless Kerr parameter)  is on the order of only $10^{-3}$ for typical neutron star parameters. This means that these birth rotation rates essentially correspond to zero angular momentum. In rare cases of pulsars with significantly faster birth spin rates (for example PSR J0534+2200 and PSR 0537-6910 with birth spin rates estimated to be $\sim 19$ ms \citep{Manchester1977} and $\sim 7$ ms \citep{Yang2025}, respectively), in what proportions the angular momentum would be transferred from the progenitor to the BM and the DM is unknown. Assuming the DM constitutes $\leq 5\%$ of the total mass of the DANS \citep{Ellis2018}, it is likely that it will possess only a fraction of the total angular momentum of the DANS. Thus, while ref.~\citep{Cipriani2025a} explores rotating DANS configurations with the DM component possessing arbitrary amounts of angular momentum (including configurations where DM counter-rotates relative to the BM), we assume DM with zero angular momentum co-rotating with the BM is a good approximation to describe rotating DANS in the accretion-powered pulsar scenario.


In this paper, we compute the structure of rapidly rotating DANS using a modification of the Rapidly Rotating Neutron Star (\texttt{RNS}) code \citep{Stergioulas1995} adapted for the two-fluid approximation and allowing for a DM component with differential rotation given by the rotating DANS' frame-dragging. This work is an extension of the code described by \cite{Konstantinou2024}, which only allowed for DM cores, to allow it to accurately compute the structure of DANS with large DM halos. In addition, we carefully examine the different definitions of mass for rotating spacetimes in general relativity in order to address the issue of quantifying the relative masses of the baryonic and dark matter in the DANS.

The rest of the paper is organized as follows. In Section~\ref{sec:model} we review how spherically symmetric DANS can be mathematically described as two fluids in hydrostatic equilibrium in general relativity, using the TOV equations. We then show how a spherically symmetric solution is used as a seed to Einstein's field equations in stationary axisymmetry to iteratively solve for rotating DANS. We also go into detail about local and global definitions of mass in general relativity, and how these two distinct definitions can introduce changes in the treatment and interpretation of mass for a two-fluid system. In Section~\ref{sec:density}, we discuss energy density distributions of baryonic and dark matter in non-rotating and rotating DANS with three characteristic types of halos. 
We compute mass-radius curves for rotating DANS in Section~\ref{sec:MR}.
Finally, in Section~\ref{sec:metric}, we show how the time-time component of non-rotating and rotating DANS with dark halos changes compared to the no-halo Schwarzschild counterpart (NHSC) approximation (introduced by \cite{Miao2022}). We use Planck units with $c = G = \hbar = 1$.

\section{Rotating DANS Model Generation}\label{sec:model}


In the two-fluid Tolman-Oppenheimer-Volkoff (TOV) \citep{Tolman1939, Oppenheimer1939} approximation, the non-gravitational interactions between BM and DM are neglected. This allows for a straightforward method for determining the baryonic surface of the NS, which is required for the analysis of X-rays emitted from the surface.  Below, we review the two-fluid formalism in spherical symmetry, and then show how the spherically symmetric solution is used as a seed to Einstein's field equations in stationary axisymmetry to iteratively solve for rotating DANS.
Two EOSs, each relating pressure and energy density for the baryonic and dark matter, are required to generate equilibrium configurations of DANS in the two-fluid approximation. 
The total energy-momentum tensor is written as
\begin{equation}
T_{T}^{\mu \nu} = T_{B}^{\mu \nu} + T_{D}^{\mu \nu} = \left(\epsilon_{B} + p_{B} \right) u_{B}^{\mu} u_{B}^{\nu} + p_{B} g^{\mu \nu} + (\epsilon_{D} + p_{D}) u_{D}^{\mu} u_{D}^{\nu} + p_{D} g^{\mu \nu},
\label{eq:energytensor}
\end{equation}
where the subscripts $B$ and $D$ represent baryonic and dark matter, respectively.

For the calculations in this paper, we use the BM-EOS, NL3$\omega \rho$L55 \citep{Horowitz2001, Pais2016}, retrieved from the CompOSE online repository of EOSs \citep{Typel2015, Oertel2017, CompOSE2022}. We model the DM as a self-interacting fermionic particle with the EOS
\citep{Narain2006,Nelson2019}
\begin{eqnarray}
\epsilon_{D} = \frac{m_{\chi}^{4}}{8 \pi^{2}} \left[ \left( 2x^{3} + x \right) \sqrt{1 + x^{2}} - \arcsinh{\left( x \right)} \right] + \frac{m_{\chi}^{4} y^{2} x^{6}}{\left( 3 \pi^{2} \right)^{2}}, \\
p_{D} = \frac{m_{\chi}^{4}}{24 \pi^{2}} \left[ \left( 2x^{3} - 3x \right) \sqrt{1 + x^{2}} + 3 \arcsinh{\left( x \right)} \right] + \frac{m_{\chi}^{4} y^{2} x^{6}}{\left( 3 \pi^{2} \right)^{2}},
\end{eqnarray}
where $m_\chi$ is the mass of the DM particle and $y$ is the self-interaction strength, defined by $y=m_{\chi}/m_I$ where $m_I$ is the mass of the self-interaction force mediator. The relativity parameter, $x$, is defined by $x=k_F/m_\chi$, where $k_F$ is the DM Fermi momentum.

\subsection{Two-fluid Formalism in Spherical Symmetry}\label{sec:modelsphere}


The metric for a spherically symmetric spacetime in Schwarzschild coordinates  is
\begin{equation}
ds^{2} = - e^{2 \Phi} dt^{2} + e^{2 \Lambda} dr^{2} + r^{2} d \theta^{2} + r^{2} \sin^{2}{\left( \theta \right)} d \phi^{2},
\label{eq:metricspher}
\end{equation}
where the metric potentials $\Phi$ and $\Lambda$ are functions of $r$.

Energy-momentum is conserved for each fluid separately $\left[ \nabla_{\mu} T_{ B \left(D \right)}^{\mu \nu} = 0 \right]$ leading to the two-fluid TOV equations \citep{Kodama1972, Sandin2009}
\begin{eqnarray}
e^{-2 \Lambda} &=& 1 - \frac{2 m_{T}}{r}, \label{eq:Lambda} \\
\frac{d \Phi}{dr} &=& \frac{ m_{T} + 4 \pi r^{3} p_{T}}{r^{2} e^{-2 \Lambda}}, \label{eq:dPhidr} \\
\frac{dp_{B}}{dr} &=& - \left( \epsilon_{B} + p_{B} \right) \frac{d \Phi}{dr}, \\
\frac{dp_{D}}{dr} &=& - \left( \epsilon_{D} + p_{D} \right) \frac{d \Phi}{dr}, \\
\frac{dm_{B}}{dr} &=& 4 \pi r^{2} \epsilon_{B}, \label{eq:dm_Bdr} \\
\frac{dm_{D}}{dr} &=& 4 \pi r^{2} \epsilon_{D}, \label{eq:dm_Ddr}
\end{eqnarray}
where $m_{T} = m_{B} + m_{D}$, $p_{T} = p_{B} + p_{D}$, and $\epsilon_{T} = \epsilon_{B} + \epsilon_{D}$. 
In these equations, the Greek and lower-case Latin symbols depend on position, although the explicit dependence is not shown.

The two-fluid TOV equations are solved using the fourth-order Runge-Kutta (RK4) method.
The equations are integrated radially outwards with the pressure and energy density of each fluid evaluated at each step. The baryonic and dark radii, $R_{B}$ and $R_{D}$, respectively, are assigned to the corresponding radial steps where the pressures vanish, i.e. $p_{B} \left( R_{B} \right) = p_{D} \left( R_{D} \right) = 0$. The TOV gravitational masses of each of the fluids are $M_{B, \textrm{TOV}} = m_{B} \left( R_{B} \right)$ and $M_{D, \textrm{TOV}} = m_{D} \left( R_{D} \right)$, with $M_{T, \textrm{TOV}} = M_{B, \textrm{TOV}} + M_{D, \textrm{TOV}}$, the total TOV gravitational mass of the DANS. The total TOV DM mass as a fraction of the total TOV mass of the DANS is then $f_{\chi, \textrm{TOV}} = M_{D, \textrm{TOV}}/M_{T, \textrm{TOV}}$.

\subsection{Two-fluid Formalism in Stationary Axisymmetry}\label{sec:modelaxi}

The metric for a stationary axisymmetric spacetime in quasi-isotropic coordinates is
\begin{equation}
ds^{2} = - e^{\gamma + \rho} dt^{2} + e^{2 \alpha} \left( d \bar{r}^{2} + \bar{r}^{2} d \theta^{2} \right) + e^{\gamma - \rho} \bar{r}^{2} \sin^{2}{\left( \theta \right)} \left( d\phi - \omega dt \right)^{2},
\label{eq:metricaxi}
\end{equation}
where $\gamma$, $\rho$, $\alpha$ and $\omega$ are functions of the quasi-isotropic radial coordinate $\bar{r}$ and $\theta$. Einstein's field equations are solved implementing the iterative Green function method proposed by \cite{Komatsu1989} and \cite{Cook1992}, using the {\tt\string RNS} implementation \citep{Stergioulas1995} modified to allow for two non-interacting fluids \citep{Konstantinou2024}.

In this spacetime, the four-velocity of the baryonic (dark) fluid is
\begin{equation}
u_{B \left( D \right)}^{\mu} = \frac{e^{- \left( \gamma + \rho \right)/2}}{\sqrt{1 - v_{B \left( D \right)}^{2}}} \left[ 1, 0, 0, \Omega_{B \left( D \right)} \right],
\label{eq:4vel}
\end{equation}
where
\begin{equation}
v_{B \left( D \right)} = \left[ \Omega_{B \left( D \right)} - \omega \right] \bar{r} \sin{\left( \theta \right)} e^{- \rho}
\label{eq:propvel}
\end{equation}
is the proper velocity of the matter with respect to a zero angular momentum observer, and $\Omega_{B \left( D \right)}$ is the angular velocity of the respective fluid as measured by an inertial observer at infinity. The specific angular momentum, $\ell_{B \left( D \right)}$ of a fluid element (ie, the angular momentum per unit mass) is defined by $\ell = u_\phi$. With the 
definition of the fluid 4-velocity in equation (\ref{eq:4vel}), the specific angular momentum of each fluid element is
\begin{equation}
    \ell_{B \left( D \right)} = u_\phi = \frac{e^{\left( \gamma - \rho \right)/2}}{\sqrt{1 - v_{B \left( D \right)}^{2}}} v_{B \left( D \right)}
    \bar{r} \sin(\theta).
\end{equation}

Consider a DANS that is initially not rotating, and thus each of its fluid components has zero angular momentum. If the DANS is in a close binary system where the companion star overflows its Roche lobe, mass will flow from the companion to the DANS. This process adds mass and angular momentum to the DANS. The simplest assumption about the companion star is that it does not contain any appreciable amount of dark matter, so that the accreted matter is purely baryonic. In the two-fluid approximation, the accreted baryonic matter can only interact with the baryonic material in the DANS, so that the accreted material only provides a torque to the baryonic component of the DANS.
As a result, the baryonic material will gain angular momentum and spin faster while the dark matter will continue to have zero angular momentum. The DANS total angular momentum will increase through the accretion process, leading to a non-zero frame-dragging metric term $\omega(\bar{r},\theta)$. 
Since the angular momentum of the DM will remain zero, the DM will acquire an angular velocity equal to the 
frame-dragging angular velocity
\begin{equation}
v_{D} = 0  \implies \Omega_{D} = \omega.
\label{eq:darkvel}
\end{equation}
Since the frame-dragging term varies with position, the dark matter will be differentially rotating, although at a rate that is much slower than the angular velocity of the baryonic matter. 

The DM differential rotation in our code is different than the standard procedure (e.g. \citep{Komatsu1989} Equation 17) using an angular momentum law with a finite parameter $A$ that is proportional to the specific angular momentum. Our prescription corresponds to $A=0$. The scenario where $\Omega \to \omega$ is described as an ``absurd" condition by \cite{Komatsu1989}, since for a single fluid this means the presence of substantial frame-dragging even in weak gravity. However, for two fluids, $\Omega_{D} = \omega$ is a physically consistent condition due to the strong gravity and rapid rotation of the baryonic fluid.

Assuming the baryonic fluid rotates rigidly in the stationary state after the spin-up process is completed, the physical scenario describes a DANS where the dark fluid follows the frame-dragging of the spacetime caused by the baryonic fluid's angular momentum.

To mathematically model this stationary state, the oblate shape of the star due to rotation is assigned by choosing a ratio of the polar and equatorial radii of the baryonic fluid in quasi-isotropic coordinates,
\begin{equation}
\bar{r}_{B \textrm{ratio}} = \bar{R}_{Bp}/\bar{R}_{Be} \leq 1.
\label{eq:rBratio}
\end{equation}
The spherically symmetric solution from the two-fluid TOV equations is used as a seed in the iterative process, and the updated value of $\bar{R}_{Bp}$ is found using
\begin{equation}
\bar{R}_{Bp} = \bar{R}_{Be}^{\prime} \bar{r}_{B \textrm{ratio}},
\label{eq:RBp}
\end{equation}
where the primed quantity represents the value of the parameter in the previous iteration.
The value of $\bar{R}_{Be}$ is then updated using
\begin{equation}
\bar{R}_{Be} = \bar{R}_{Be}^{\prime} \sqrt{\frac{2 \left( h_{Bc} - h_{Bp} \right)}{\bar{\gamma}_{Bp}^{\prime} + \bar{\rho}_{Bp}^{\prime} - \bar{\gamma}_{c}^{\prime} - \bar{\rho}_{c}^{\prime}}},
\label{eq:RBe}
\end{equation}
where $h$ is the enthalpy, and the subscripts $c$ and $Bp$ represent values of the quantities evaluated at the centre of the star, and at $\bar{R}_{Bp}$, respectively \citep{Cook1994}. The equatorial and polar radii of the dark fluid, $\bar{R}_{De}$ and $\bar{R}_{Dp}$, are assigned to the radial coordinates where the enthalpy of the dark fluid, $h_{D}$, vanishes. The dark enthalpy, $h_{D}$, is calculated using Equation A25 of \cite{Cook1994} with $A = 0$ and $v = 0$. Note that this step is different from that of \cite{Konstantinou2024} where an $\bar{r}_{D \textrm{ratio}} = \bar{R}_{Dp}/\bar{R}_{De} = 1$ was chosen, which introduces a non-zero angular momentum to the dark fluid.

The iterative process is continued until $\left| \bar{R}_{Be} - \bar{R}_{Be}^{\prime} \right|/\bar{R}_{Be} < 10^{-5}$, which defines the precision of the code. Finally, the radii of the DANS in Schwarzschild coordinates, $R$, is recovered using
\begin{equation}
R = e^{\left( \gamma - \rho \right)/2} \bar{R}.
\label{eq:R}
\end{equation}

While converging to a solution, in each iteration, the {\tt\string RNS} code computes the entire spacetime where the radial coordinate extends from zero to infinity. To make this problem numerically tractable, auxiliary variables
\cite{Cook1992}
\begin{equation}
s = \frac{\bar{r}}{\bar{R}_{Be} + \bar{r}}
\label{eq:s}
\end{equation}
and
\begin{equation}
\mu = \cos{\left( \theta \right)}
\label{eq:mu}
\end{equation}
are introduced, where $0 \leq s < 1$ for $0 \leq \bar{r} < \infty$, and $0 \leq \mu \leq 1$ for $\pi/2 \geq \theta \geq 0$. All the DANS computed in this paper using the two-fluid {\tt\string RNS} code use an $s \times \mu$ grid size of $2000 \times 400$, which is a much higher spatial resolution than is typically used in the computation of stationary neutron star models without dark matter. The choice of $\bar{R}_{Be}$ as the scaling factor in Equation~\eqref{eq:s} is important to ensure a DANS is sufficiently resolved up to the baryonic radius, and the entire DANS is accurately computed. For reference, the DANS with the largest value of $R_{De}$ in Table \ref{tab:densityDANS} $\left( 530.12 \textrm{ km} \right)$ has a value of $s = 0.978402$ at $R_{De}$.
 
\subsection{Definitions of Mass}\label{sec:modelmass}

In general relativity, it is not normally possible to define a local energy for the gravitational field, since this would violate the equivalence principle. As a result, there is no general definition of the total energy at a point of spacetime. In spherical symmetry, however, one can define a total mass-energy, $m_T(r)$ found within a sphere of radius $r$ through the differential equations \eqref{eq:dm_Bdr} and \eqref{eq:dm_Ddr}. Integrating these two equations over all space leads to the mathematical definition of the TOV mass of each fluid component
\begin{equation}
M_{B \left( D \right), \textrm{TOV}} = 4 \pi \int_{0}^{\infty} dr r^{2} \epsilon_{B \left( D \right)}.
\label{eq:MXTOV}
\end{equation}
The sum $M_{TOV} = M_{B,TOV} + M_{D,TOV}$ is an invariant, physical observable. However, the separate masses $M_{B(D),TOV}$ are not measurable quantities, although they are often used to separately characterize the gravitational masses of the baryonic and DM components of non-rotating DANS. The problems related to the definitions of mass for two-fluid systems in general relativity are commented on by \cite{Cipriani2025a}. Since the definition of mass is so important for the discussion of measurable quantities for DANS with halos, we explore the differences in mass definitions in much more detail in this section. 

In a general stationary axisymmetric spacetime, there is no local definition of the energy. However, there is a global definition based on an integral at spatial infinity due to \cite{1959PhRv..113..934Komar}. When Stokes' theorem is applied, the Komar mass for each fluid component corresponds to a three-dimensional integral over all space defined by
\citep{2013rrs..book.....Friedman}
\begin{equation}
M_{B \left( D \right), \textrm{g}} = \int_{0}^{2 \pi} d \phi \int_{0}^{\pi} d \theta \int_{0}^{\infty} dr \sqrt{-g} \left[ -2 T_{0 B \left( D \right)}^{0} + T_{\mu B \left( D \right)}^{\mu} \right],
\label{eq:MXglob}
\end{equation}
where $g$ is the determinant of the metric tensor. These mathematical definitions of the global baryonic and dark masses are possible in the two-fluid formalism due to the linearity of the stress-energy tensor, however, only the sum $M_{T,g} = M_{B,g} + M_{D,g}$ has an invariant physical meaning. For stationary axisymmetric spacetimes, $M_{T,g}$ is also equivalent to the ADM mass.

\subsubsection{Definitions of Global and Local Mass in Spherical Symmetry}
\label{sec:DefMassSS}

In spherical symmetry, the global definition of mass \eqref{eq:MXglob} for each fluid
takes the form
\begin{equation}
M_{B \left( D \right), \textrm{g}} = 4 \pi \int_{0}^{\infty} dr r^{2} e^{\Phi + \Lambda} \left[ \epsilon_{B \left( D \right)} + 3 p_{B \left( D \right)} \right] \neq M_{B \left( D \right), \textrm{TOV}}.
\label{eq:MXglobsphere}
\end{equation}
While the local and global definitions of mass for the individual fluid components do not agree, the sums of the components do satisfy
\begin{equation}
M_{T, \textrm{g}} = M_{T, \textrm{TOV}}.
\label{eq:MT}
\end{equation}
The equality of the two definitions in spherical symmetry can be proved through integration by parts and application of the two-fluid TOV equations in the same way that the equality is proved for a single fluid (e.g., \cite{1975pbrg.book.....Lightman}).
The same process of integration by parts  and utilizing the two-fluid TOV equations can be used to show that the local and global definitions of baryonic mass (or dark matter mass) 
differ by an integral
\begin{align}
\nonumber M_{B \left( D \right), \textrm{g}} = M_{B \left( D \right), \textrm{TOV}} + 4 \pi \int_{0}^{\infty} dr r e^{\Phi + 3 \Lambda} \big \{ &m_{T} \left[  \epsilon_{B \left( D \right)} + p_{B \left( D \right)} \right] - m_{B \left( D \right)} \left[  \epsilon_{T} + p_{T} \right] \\
&+ 4 \pi r^{3} \left[ \epsilon_{B \left( D \right)} p_{T} - p_{B \left( D \right)} \epsilon_{T} \right] \big \},
\label{eq:MXglobsphereparts}
\end{align}
where we can make use of the  relation that $e^{\Phi+\Lambda} = 1$ when evaluated at $r>R_D$ where the pressure vanishes.

The integral defining the differences between global and local component masses can be simplified to 
\begin{align}
    \nonumber M_{B,\textrm{g}} - M_{B,TOV} &= - (M_{D,\textrm{g}} - M_{D,TOV})\\
    &= 4 \pi \int_{0}^{\infty} dr r e^{\Phi + 3 \Lambda} \big \{
m_D(\epsilon_B + p_B) - m_B(\epsilon_D+ p_D)
+ 4 \pi r^3(\epsilon_B p_D - p_B\epsilon_D)
    \big \}.
    \label{eq:massdiff}
\end{align}
The mixing of terms between the baryonic and dark fluids in Equations \eqref{eq:MXglobsphereparts} and \eqref{eq:massdiff} shows that $M_{B \left( D \right), \textrm{g}}$ includes contributions from both fluids. So, while masses (global or local) can be calculated for each of the fluids separately, their interpretation as the separate gravitational masses of each of the fluids should not be taken literally.

Since we are considering DANS with DM energy densities that are typically 1 - 3 orders of magnitude smaller than the baryonic energy densities, the asymmetries present in the 
integral in Equation~\eqref{eq:massdiff} result in a non-zero value of the integral. 

\subsubsection{Definition of Mass Within a Closed Finite Surface in Spherical Symmetry}

Refs. \cite{Miao2022} and \cite{Shawqi2024} showed that in spherical symmetry, the enclosed TOV mass within the baryonic radius $m_{T} \left( R_{B} \right)$ is an important quantity relevant to mass measured using the gravitational lensing method (as is done analysing X-rays observed by telescopes such as NICER and Chandra) for stars with DM halos. A non-local version of such a mass can be defined by truncating the integral appearing in the global mass definition at the baryonic radius, $R_{B}$. When integrating by parts the outer surface is now the baryonic surface (instead of infinity), where the DM pressure does not vanish and the quantity $e^{\Phi \left( R_{B} \right) + \Lambda \left( R_{B} \right)}$ is not equal to one.
This results in
\begin{align}
M_{T, \textrm{g}} \left( R_{B} \right) &= 4 \pi \int_{0}^{R_{B}} dr r^{2} e^{\Phi + \Lambda} \left( \epsilon_{T} + 3 p_{T} \right) = e^{\Phi \left( R_{B} \right) + \Lambda \left( R_{B} \right)} \left[ m_{T} \left( R_{B} \right) + 4 \pi R_{B}^{3} p_{D} \left( R_{B} \right) \right] \label{eq:MT(RB)globsphere}\\
&\neq m_{T} \left( R_{B} \right), \label{eq:mT(RB)}
\end{align}
where the local and non-local definitions of the total mass within the baryonic surface agree only if the region exterior to the baryonic surface is in vacuum. 

In a similar way, the baryonic or DM mass inside the baryonic surface 
can be defined using the global mass definition,
\begin{align}
\nonumber M_{B \left( D \right), \textrm{g}} \left( R_{B} \right) &= && \mkern-58mu e^{\Phi \left( R_{B} \right) + \Lambda \left( R_{B} \right)} \left[ m_{B \left( D \right)} \left( R_{B} \right) + 4 \pi R_{B}^{3} p_{B \left( D \right)} \left( R_{B} \right) \right] \\
\nonumber & && \mkern-58mu+ 4 \pi \int_{0}^{R_{B}} dr r e^{\Phi + 3 \Lambda} \big \{m_{T} \left[  \epsilon_{B \left( D \right)} + p_{B \left( D \right)} \right] - m_{B \left( D \right)} \left[  \epsilon_{T} + p_{T} \right] \\
& && \mkern110mu+ 4 \pi r^{3} \left[ \epsilon_{B \left( D \right)} p_{T} - p_{B \left( D \right)} \epsilon_{T} \right] \big \} \label{eq:MX(RB)globsphereparts}\\
&\neq && \mkern-58mu m_{B \left( D \right)} \left( R_{B} \right), \label{eq:mX(RB)}
\end{align}
which also does not agree with the local definitions of baryonic or DM mass inside of $R_B$. 

The gravitational mass felt by a test particle is the total energy of the system, including the negative gravitational binding energy. Since both fluids act as a source of the gravitational field in a non-linear way, one can't cleanly divide up the gravitational field (or the binding energy) into components caused by the dark matter or the baryonic matter separately. 

\subsubsection{Definitions of the Dark Matter Cloud Mass in Spherical Symmetry}

The DM cloud mass is defined as the difference between the mass enclosed between the DM and baryonic surfaces. 
The definition of the cloud mass using the TOV mass, $M_{c, \textrm{TOV}}$, is 
\begin{equation}
    M_{c, \textrm{TOV}} = 
    M_{T, \textrm{TOV}} - m_{T} \left( R_{B} \right).
\end{equation}
The TOV  baryonic mass, is defined by $M_{B, \textrm{TOV}} = m_{B} \left( R_{B} \right)$, so the TOV cloud mass is just the difference between the DM TOV mass,
and the DM located inside the baryonic surface,
\begin{equation}
M_{c, \textrm{TOV}} = M_{D, \textrm{TOV}} - m_{D} \left( R_{B} \right).
\label{eq:McTOVD}
\end{equation}

The non-local cloud mass, $M_{c,g}$, defined through the difference between the global mass and the non-local mass evaluated at the baryonic surface is 
\begin{align}
M_{c, \textrm{g}} &= M_{T, \textrm{g}} - M_{T, \textrm{g}} \left( R_{B} \right) = M_{T, \textrm{TOV}} - e^{\Phi \left( R_{B} \right) + \Lambda \left( R_{B} \right)} \left[ m_{T} \left( R_{B} \right) + 4 \pi R_{B}^{3} p_{D} \left( R_{B} \right) \right] \label{eq:McglobT}\\
&\neq M_{T, \textrm{TOV}} - m_{T} \left( R_{B} \right) = M_{c, \textrm{TOV}} \label{eq:McTOVT},
\end{align}
which does not agree with the local definition of the cloud mass.


The non-local cloud mass can't be 
defined in terms of differences of only DM masses since
\begin{equation}
M_{B, \textrm{g}} - M_{B, \textrm{g}} \left( R_{B} \right) = M_{B, \textrm{TOV}} \left[ 1 - e^{\Phi \left( R_{B} \right) + \Lambda \left( R_{B} \right)} - 4 \pi \int_{R_{B}}^{R_{D}} dr r e^{\Phi + 3 \Lambda} \left( \epsilon_{D} + p_{D} \right) \right] \neq 0, \label{eq:MBglob-MB(RB)glob} 
\end{equation}
which means that some of the baryonic mass in the global definition is ``located" in the cloud region.

As a result, the non-local cloud mass is not equivalent to the difference between the global DM mass evaluated at infinity and at the baryonic surface since
\begin{align}
M_{c, \textrm{g}} &\neq && \mkern-120mu M_{D, \textrm{g}} - M_{D, \textrm{g}} \left( R_{B} \right) \label{eq:McglobD0}\\
\nonumber &= && \mkern-120mu M_{D, \textrm{TOV}} - e^{\Phi \left( R_{B} \right) + \Lambda \left( R_{B} \right)} \left[ m_{D} \left( R_{B} \right) + 4 \pi R_{B}^{3} p_{D} \left( R_{B} \right) \right] \\
& && \mkern-120mu + 4 \pi M_{B, \textrm{TOV}} \int_{R_{B}}^{R_{D}} dr r e^{\Phi + 3 \Lambda} \left( \epsilon_{D} + p_{D} \right). \label{eq:McglobD1}
\end{align}
This further signifies that it is not appropriate to interpret Equation~\eqref{eq:MXglobsphereparts} as the gravitational mass of each of the fluids separately. In the global definition of mass, it is more appropriate to consider the total mass of both fluids together.

\subsubsection{Definition of Mass in Axisymmetry}

There is no local definition of mass in a spacetime with angular momentum, so the only physically meaningful definition of mass is the global definition. Only the total mass $M_g$ can be measured. The mathematical split of the global mass into baryonic and DM components in \eqref{eq:MXglob} for rotating DANS is
\begin{align}
\nonumber M_{B \left( D \right), \textrm{g}} = 4 \pi \int_{0}^{\infty} d\bar{r} \bar{r}^{2} \int_{0}^{\pi/2} d\theta \sin{\left( \theta \right)} e^{2 \alpha + \gamma} \Bigg \{ & \frac{ \left[ \epsilon_{B \left( D \right)} + p_{B \left( D \right)} \right]}{1 - v_{B \left( D \right)}^{2}} \\
\nonumber & \times \left[ 1 + v_{B \left( D \right)}^{2} + 2 \omega \bar{r} v_{B \left( D \right)} e^{-\rho} \sin{\left( \theta \right)} \right] \\
& + 2 p_{B \left( D \right)} \Bigg \}.
\label{eq:MXglobaxi}
\end{align}
Since we are only considering dark matter halos that have zero angular momentum,  $v_{D} = 0$, 
and the DM component of the global mass is
\begin{equation}
M_{D, \textrm{g}} = 4 \pi \int_{0}^{\infty} d\bar{r} \bar{r}^{2} \int_{0}^{\pi/2} d\theta \sin{\left( \theta \right)} e^{2 \alpha + \gamma} \left( \epsilon_{D} + 3 p_{D} \right).
\label{eq:MDglobaxiv0}
\end{equation}

Since in the limit of zero rotation, Equations \eqref{eq:MXglobaxi} and \eqref{eq:MDglobaxiv0} reduce to  Equation~\eqref{eq:MXglobsphereparts}, there is mixing of masses between the baryonic and dark fluids in axisymmetry, similar to what is seen in the case of spherical symmetry. Thus, the separate global mass of each fluid should not be literally interpreted as the separate gravitational mass of each fluid, both in spherical symmetry and in axisymmetry.

\subsection{Comparison between Two-Fluid TOV and {\tt\string RNS} in Spherical Symmetry}\label{sec:modelcompare}

Comparisons of non-rotating DANS computed using the two-fluid version of 2D {\tt\string RNS} code with the results of the 1D-TOV equations serve as useful code accuracy tests. In addition, since we are interested in exploring the differences between the local and global definitions of mass, it is important to know the size of our numerical errors in order to ensure that we are resolving real differences introduced by the definition of the global mass. 

Since the 1D TOV equations are more accurate than the 2D code, relative differences in the values of the quantities $R_B$, $R_D$, and $M_{T}$ that should be the same in both codes serve as accuracy tests of the two-fluid {\tt\string RNS} code. We will also compare the differences between the DM and BM local and global mass definitions for the spherically symmetric solutions. 


Spherically symmetric solutions in {\tt\string RNS} are found by setting $\bar{r}_{B\textrm{ratio}} = 1$. Since the baryonic fluid does not cause any frame dragging of the spacetime with this choice of $\bar{r}_{B\textrm{ratio}}$, fixing $v_{D} = 0$ ensures the dark fluid is also spherically symmetric. The two-fluid TOV results that we use for comparison are in fact, the seed solution to the iterative process that {\tt\string RNS} uses.

We compare DM halos produced from three combinations of $m_{\chi}$ and $y$, and a fixed $f_{\chi, \textrm{TOV}} = 0.05$ \citep{Ellis2018}. The combinations are (i) $m_{\chi} = 0.3$ GeV, $y = 0$, (ii) $m_{\chi} = 1$ GeV, $y = 100$, and (iii) $m_{\chi} = 1$ GeV, $y = 1000$. These are the same models computed by \cite{Shawqi2024} using the two-fluid TOV equations. For a DANS with $M_{T, \textrm{TOV}} = 1.4 M_{\odot}$ and $f_{\chi, \textrm{TOV}} = 0.05$ (i.e. $M_{D, \textrm{TOV}} = 0.07 M_{\odot}$), these combinations of $m_{\chi}$ and $y$ produce DM halos of three characteristic types. Using the terminology introduced by \cite{Shawqi2024}, (i) produces a \textit{compact} halo with $M_{c, \textrm{TOV}}/M_{D, \textrm{TOV}} = 0.124$, (ii) produces an \textit{intermediate} halo with $M_{c, \textrm{TOV}}/M_{D, \textrm{TOV}} = 0.793$, and (iii) produces a \textit{diffuse} halo with $M_{c, \textrm{TOV}}/M_{D, \textrm{TOV}} = 0.997$. Note that (i) produces DM cores for $M_{T, \textrm{TOV}} \gtrsim 2.36 M_{\odot}$, but for convenience we continue to refer to them as \textit{compact} DANS. We adopt the choices of $m_{\chi}$, $y$ and $f_{\chi}$, made by ref.~\cite{Shawqi2024} which produce DM \textit{halos} of the characteristically different types. Since these are the three types of halos that can form, we only require one example of parameters that produce each.  
This is not an exhaustive sampling of DM parameters, and it could be of interest to explore the parameter space more thoroughly in the future. This also allows for relevant comparisons with the non-rotating stars computed by ref.~\cite{Shawqi2024}.

In this section, we compute fractional difference comparisons for various quantities for DANS in spherical symmetry, calculated by solving the two-fluid TOV equations and the {\tt\string RNS} code
for DANS with the same values of central BM and DM densities. The fractional difference is 
defined by 
\begin{equation}
\Delta_{\textrm{TOV}} Q = \frac{\left| Q_{{\tt\string RNS}} - Q_{\textrm{TOV}} \right|}{Q_{\textrm{TOV}}},
\label{eq:DeltaTOV}
\end{equation}
where $Q_{\textrm{TOV}}$ represents quantities obtained by solving the two-fluid TOV equations and $Q_{{\tt\string RNS}}$ refers to quantities produced by the {\tt\string RNS} code. If the quantity being compared is a mass, the mass produced by the {\tt\string RNS} code is computed using the global definition discussed in Section~\ref{sec:modelmass}, i.e. $M_{{\tt\string RNS}} = M_{\textrm{g}}$. Along with DANS with three combinations of $m_{\chi}$ and $y$, results for pure baryonic NSs with no DM are also compared for appropriate quantities. We show results for a range of central densities, corresponding to DANS with total masses between $M_{T, \textrm{TOV}} = 1 M_{\odot}$ and maximum $M_{T, \textrm{TOV}}$.

\subsubsection{Code Accuracy Tests for Spherical Symmetry}

\begin{figure}[ht!]
\includegraphics[width=\textwidth]{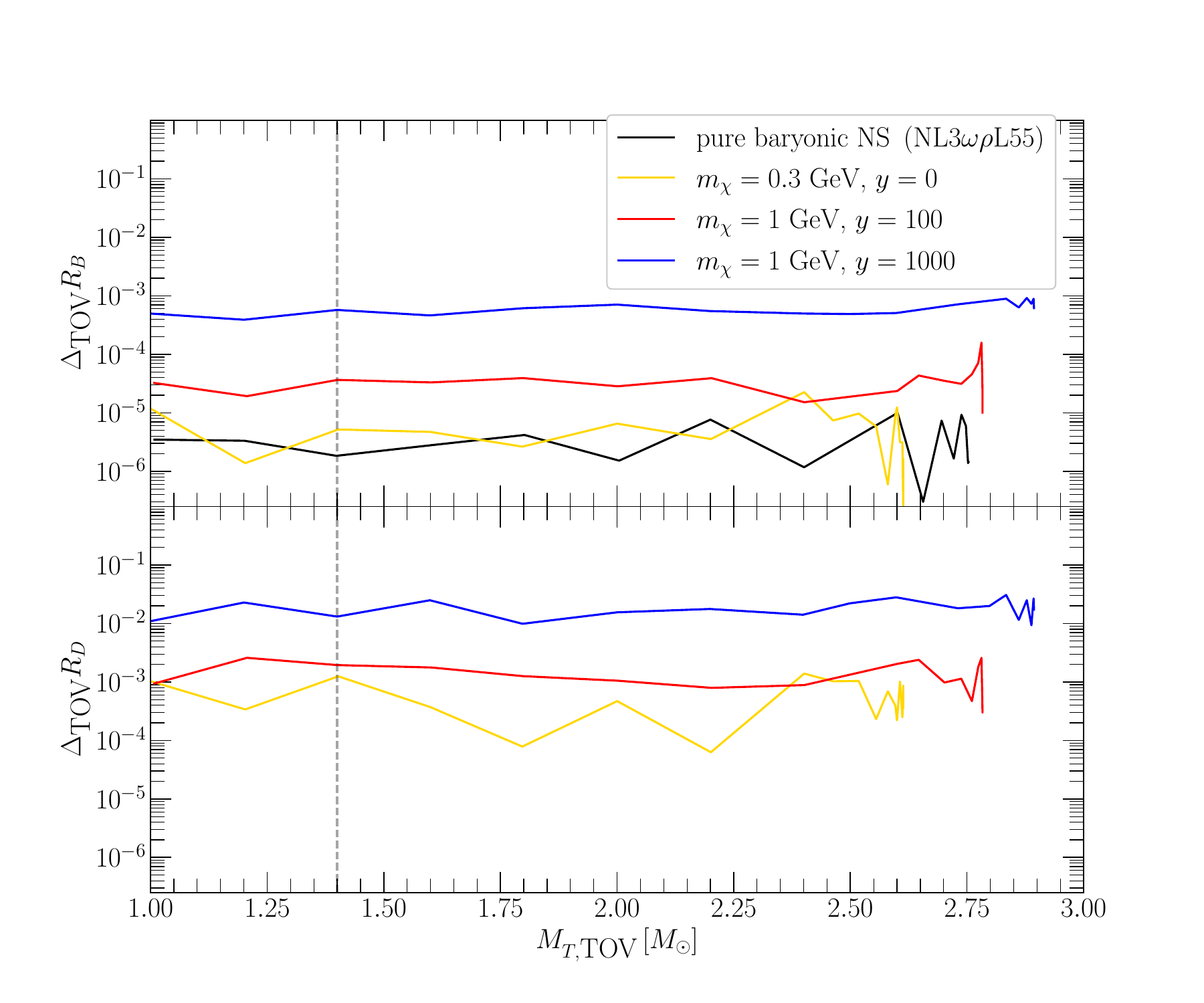}
\caption{{{Fractional differences in $R_{B}$ (upper) and  $R_{D}$ (lower), calculated in spherical symmetry using the {\tt\string RNS} code and by solving the two-fluid TOV equations. The differences arise from numerical resolution errors associated with the two-fluid {\tt\string RNS} code. The dashed vertical line represents the $M_{T, \textrm{TOV}} = 1.4 M_{\odot}$ stars shown in Table \ref{tab:densityDANS} and Figures ~\ref{fig:density} and~\ref{fig:omega}.}}
\label{fig:TOVRNScompareR}}
\end{figure}

The accuracy of the two-fluid {\tt\string RNS} code is tested through the fractional errors in $R_B$, $R_D$, and $M_T$, since these quantities are independent of the calculation method. Figure~\ref{fig:TOVRNScompareR} (upper panel) compares the fractional differences in baryonic radius $R_{B}$ obtained using the two methods, $\Delta_{\textrm{TOV}} R_{B}$, as a function of $M_{T, \textrm{TOV}}$. For the pure baryonic NSs (shown in black), the differences are $\lesssim 10^{-5}$ and represent the numerical resolution of radius calculations for the standard single-fluid {\tt\string RNS} code. 
These fractional errors are of a similar order of magnitude \citep{1998A&AS..132..431Nozawa} to the fractional differences for non-rotating stars found through comparisons between  {\tt\string RNS} and the spectral methods based {\tt\string BGSM} code \citep{1993A&A...278..421Bonazzola}. The addition of a dark matter halo reduces the accuracy for finding the baryonic radius, as can be seen in the coloured curves shown in Figure~\ref{fig:TOVRNScompareR} (upper panel).
The differences for the compact halos (yellow) are of a similar order as the numerical resolution. The intermediate halos have $\Delta_{\textrm{TOV}} R_{B} \lesssim 10^{-4}$. The largest differences of $\lesssim 10^{-3}$ are obtained for the diffuse halos. For all halos, there is no clear dependence on $M_{T, \textrm{TOV}}$ for the range of masses considered. Our code has errors in finding the location of the baryonic radius that increase as the size of the dark matter halo increases. However, since the errors are less than 0.1\% for the largest halos, this accuracy is sufficient for understanding the properties of these DANS. In particular, note that the values of $R_B$ for the blue and red models shown in Table \ref{tab:densityDANS} differ by a fractional difference that is an order of magnitude larger than $\Delta_{\textrm{TOV}} R_{B}$ for the blue model, so the change in $R_B$ arising from changing the dark matter properties is real (not a numerical artifact).

Figure~\ref{fig:TOVRNScompareR} (lower panel) shows the errors in the dark radius, $\Delta_{\textrm{TOV}} R_{D}$, as a function of $M_{T, \textrm{TOV}}$. The numerical errors in $R_D$ are larger than $R_B$, since our method prioritizes accuracy in finding the baryonic radius. Similar to the differences in $R_{B}$, the compact halos have the lowest fractional differences of $\sim \mathcal{O} \left( 10^{-3} \right)$ for $R_{D}$ obtained using the two methods, intermediate halos have differences $\sim \mathcal{O} \left( 10^{-3} \right)$, and the diffuse halos have differences $\sim \mathcal{O} \left( 10^{-2} \right)$. For the masses considered, the compact stars have $R_{D, \textrm{TOV}} \sim 10.7 \textrm{\textendash} 23.0$ km, the intermediate halos have $R_{D, \textrm{TOV}} \sim 34.9 \textrm{\textendash} 50.2$ km, and the diffuse halos have $R_{D, \textrm{TOV}} \sim 511 \textrm{\textendash} 525$ km (for all the DANS, $R_{D, \textrm{TOV}}$ decreases with increasing  $M_{T, \textrm{TOV}}$). Thus, for both $R_{B}$ and $R_{D}$, the stars with the largest halos have the largest relative differences. Again for all cases, there is no clear dependence on $M_{T, \textrm{TOV}}$. In {\tt\string RNS}, the same spatial grid size is used to compute all DANS, irrespective of the size of the halos. As a result, the numerical resolution worsens for the larger halos, leading to larger numerical errors when determining the outer boundaries of the dark halos.  

\begin{figure}[ht!]
\centering
\includegraphics[width=0.641\textwidth]{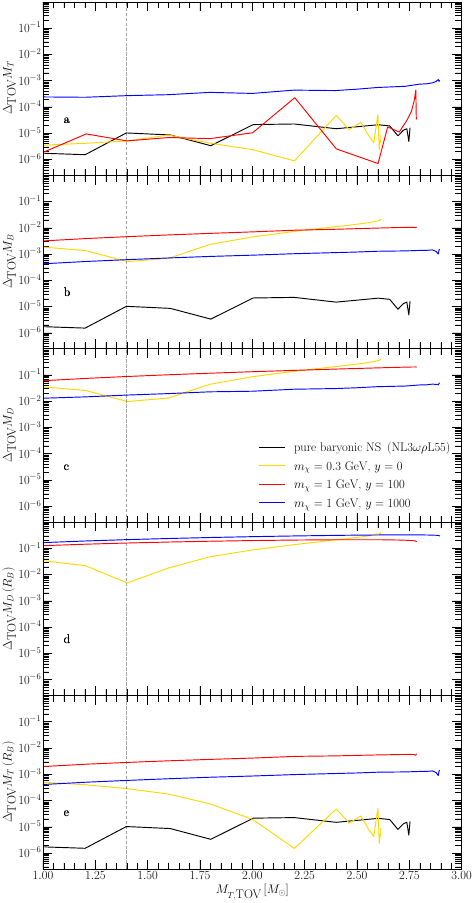}
\caption{{{Fractional differences in five different mass definitions calculated in spherical symmetry using the {\tt\string RNS} code and by solving the two-fluid TOV equations.
The differences in $M_{T}$ arise from numerical resolution and errors associated with the two different methods of calculation. Differences in $M_{B}$, $M_{D}$, $M_{D} \left( R_{B} \right)$ and $M_{T} \left( R_{B} \right)$ arise from the difference in the definitions of global and local (TOV) masses. Dashed vertical line represents the $M_{T, \textrm{TOV}} = 1.4 M_{\odot}$ stars shown in Table \ref{tab:densityDANS} and Figures \ref{fig:density} and~\ref{fig:omega}.}}
\label{fig:TOVRNScompareM}}
\end{figure}

Figure~\ref{fig:TOVRNScompareM}a shows the fractional difference in the total mass, $\Delta_{\textrm{TOV}} M_{T}$, as a function of $M_{T, \textrm{TOV}}$. According to Equation~\eqref{eq:MT}, the two definitions of mass agree exactly on $M_{T}$, so the differences between $M_{T, \textrm{TOV}}$ and $M_{T, {\tt\string RNS}}$ are due to the numerical error in computing the mass integral in the two-fluid {\tt\string RNS} code. 

The numerical resolution of the mass is similar to the resolution of the radius for the cases of 
 the pure baryonic NSs and the compact DANS. This is also generally true for the intermediate halos, but with some fluctuations for $M_{T, \textrm{TOV}} \gtrsim 2 M_{\odot}$. The differences for the diffuse halos are of $\sim \mathcal{O} \left( 10^{-4} \right)$. The cause of the larger size of differences for the diffuse halos and for some intermediate halos is the numerical resolution issue that occurs for the computation of the dark radius.

\subsubsection{Difference Between Local and Global Mass Definitions}

As shown in Section~\ref{sec:DefMassSS}, the local and global definitions of the baryonic (or dark matter) mass in a star do not agree, even in the case of spherical symmetry. In all past work restricted to non-rotating stars, mass content has been quantified in terms of the local mass definitions. However, when stars rotate, only the global mass definition is available. In this section, we explore the typical magnitudes of the differences in mass definitions for spherical symmetry so that we can understand the numerical results for rotating DANS.

Figure~\ref{fig:TOVRNScompareM}b shows the fractional difference between the local and global masses of the baryonic mass of the star,  $\Delta_{\textrm{TOV}} M_{B}$. For the pure baryonic NSs, the differences are $\sim \mathcal{O} \left( 10^{-6} \right)$, which sets the numerical resolution for mass computations. For all three types of halos, the fractional difference in the baryonic mass is 2-3 orders of magnitude larger than the numerical resolution limit of the code, showing that these are real differences, although they are very small. 
In contrast to the radius comparisons, the diffuse halos have the lowest differences of $\sim \mathcal{O} \left( 10^{-4} \right)$. The compact halos have lower differences than the intermediate halos for $M_{T, \textrm{TOV}} \lesssim 2.2 M_{\odot}$, and both have similar differences of $\sim \mathcal{O} \left( 10^{-3} \right)$ for  $M_{T, \textrm{TOV}} \gtrsim 2.2 M_{\odot}$. For all cases, differences tend to increase with increasing $M_{T, \textrm{TOV}}$, corresponding to DANS with larger gravitational fields. {{Although we do not explicitly show the signs of the differences in Figure~\ref{fig:TOVRNScompareM}, we find that $M_{B, \textrm{TOV}} > M_{B, {\tt\string RNS}}$ for the diffuse and intermediate halos. For the DANS with compact halos, $M_{B, \textrm{TOV}} > M_{B, {\tt\string RNS}}$ for $M_{T, \textrm{TOV}} \lesssim 1.5 M_{\odot}$, and $M_{B, \textrm{TOV}} < M_{B, {\tt\string RNS}}$ for $M_{T, \textrm{TOV}} \gtrsim 1.5 M_{\odot}$.}}

The absolute difference in the DM mass of the DANS is simply the negative of the absolute difference of the baryonic mass of the DANS, since the total masses are the same. However, since we are only considering DANS with small amounts of dark matter, the fractional difference in the DM mass is much larger than in the case of the baryonic mass difference. Figure~\ref{fig:TOVRNScompareM}c shows the fractional differences in $M_{D}$. 
The diffuse halos have the lowest differences of $\sim \mathcal{O} \left( 10^{-2} \right)$. The compact halos have lower differences than the intermediate halos for $M_{T, \textrm{TOV}} \lesssim 2.2 M_{\odot}$, and both have similar differences of $\sim \mathcal{O} \left( 10^{-1} \right)$ for  $M_{T, \textrm{TOV}} \gtrsim 2.2 M_{\odot}$. Again, for all cases, differences tend to increase with increasing $M_{T, \textrm{TOV}}$. The differences between the two methods of calculating mass are most pronounced for the calculation of $M_{D}$. 
The differences in both $M_{B}$ and $M_{D}$ between the two mass definitions arise from the size of the integral in Equation~\eqref{eq:massdiff}.

In our calculations of DANS with $f_{\chi,TOV} = 0.05$ shown in Figure~\ref{fig:TOVRNScompareM} the fractional difference in the dark mass ranges from 1\% to 10\% between the local and global mass definitions. This means that a dark matter mass fraction defined in terms of $M_{D,g}$ instead of $M_{D,TOV}$ will differ by up to 10\% from the definition in terms of the TOV dark mass. Since the dark mass fraction is small to begin with, this represents a fairly small change. 


Next, in Figure~\ref{fig:TOVRNScompareM}d we show values of $\Delta_{\textrm{TOV}} M_{D} \left( R_{B} \right)$ where the integrals in the mass calculations are truncated at $r = R_{B}$. The sizes of the differences in $M_{D} \left( R_{B} \right)$ are comparable to those of $M_{D}$. Overall, the DANS with compact halos have the lowest difference of $\sim \mathcal{O} \left( 10^{-2} \right)$. The intermediate and diffuse halos have differences of $\sim \mathcal{O} \left( 10^{-1} \right)$. The size of differences for the compact stars rises with increasing $M_{T, \textrm{TOV}}$ and is similar to those of the intermediate and diffuse halos for high masses. The size of the differences for all DANS arises from the size of the differences between the quantities in equations \eqref{eq:MX(RB)globsphereparts} and \eqref{eq:mX(RB)}.

Finally, in Figure~\ref{fig:TOVRNScompareM}e we show the fractional differences between calculating $M_{T, \textrm{g}} \left( R_{B} \right)$ using the global definition and $m_{T} \left( R_{B} \right)$ using the local (TOV) definition. The differences for the pure baryonic NSs are the same as those for $M_{B}$ and $M_{T}$, and represent the numerical resolution for mass calculation using the two methods. The compact halos have differences of $\sim \mathcal{O} \left( 10^{-4} \right)$ near $M_{T, \textrm{TOV}} \sim 1 M_{\odot}$, but fall to values comparable to numerical resolution for higher masses. This is because as $R_{D}$ decreases with increasing masses for the compact stars, the dark halos transition to dark cores and thus, $M_{T, \textrm{g}} \left( R_{B} \right)$ gets closer and closer to $M_{T, \textrm{g}} = M_{T, \textrm{TOV}}$. Both the intermediate and diffuse halos have differences of $\sim \mathcal{O} \left( 10^{-3} \right)$, with the intermediate values having larger differences for all values of $M_{T, \textrm{TOV}}$ considered.


Overall, the differences in $M_{B}$, $M_{D}$, $M_{D} \left( R_{B} \right)$ and $M_{T} \left( R_{B} \right)$ arise from the difference in the definitions of global and local (TOV) masses and are numerically resolvable by our code. 

\section{Distributions of Energy Density and Frame-dragging of Spacetime} \label{sec:density}

\begin{figure}[ht!]
\includegraphics[width=\textwidth]{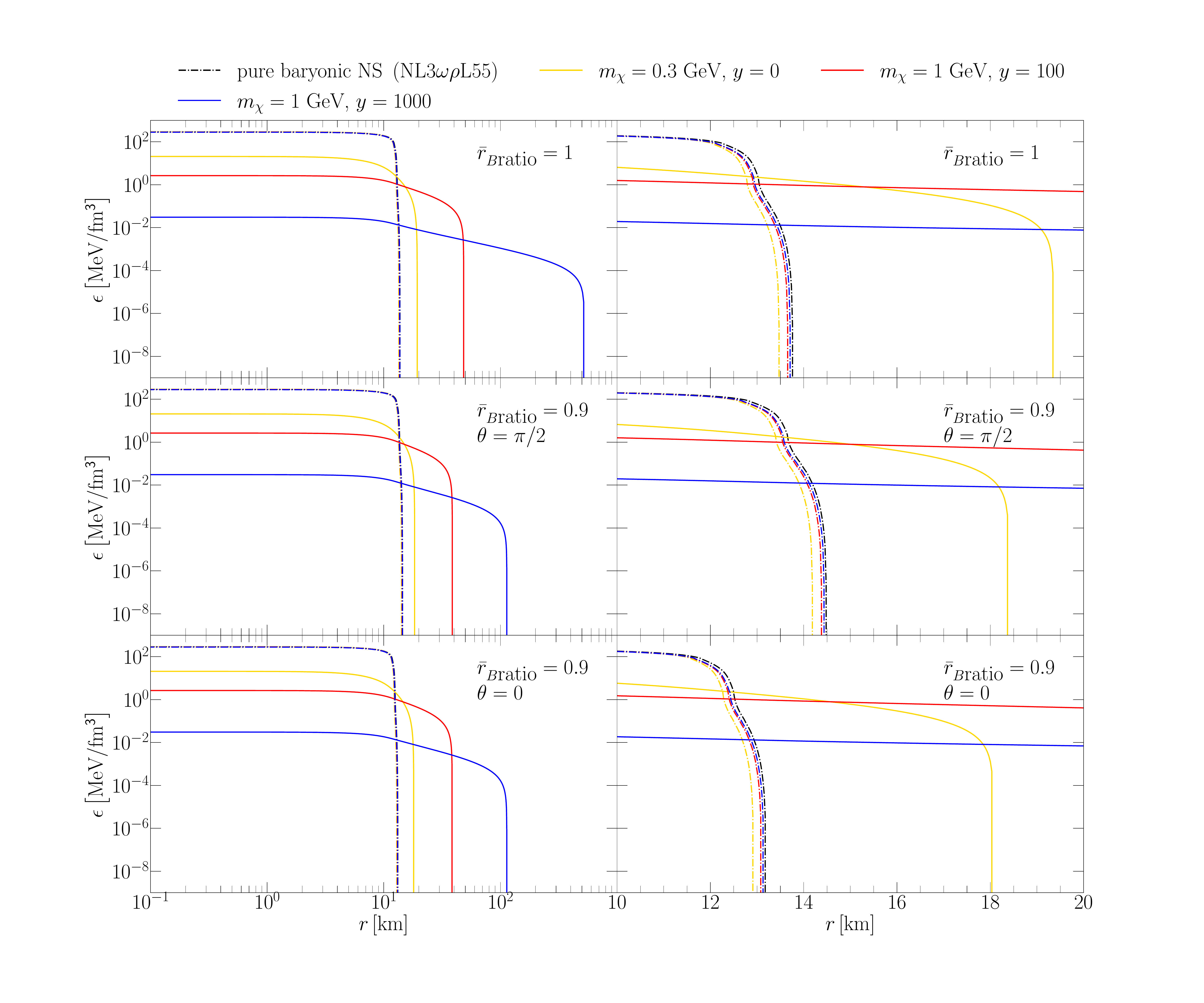}
\caption{Energy density distributions as a function of radial distance from the centre for DANS in Table \ref{tab:densityDANS}. Top: Baryonic (dash-dot) and dark halo (solid) energy densities of $M_{T} = 1.4 M_{\odot}$ DANS models with $f_{\chi, \textrm{TOV}} = 0.05$ and $\bar{r}_{B \textrm{ratio}} = 1$. Center: Equatorial energy densities of rotating ($\bar{r}_{B \textrm{ratio}} = 0.9$) DANS models with the same central energy densities as the models in the top plots. Bottom: Polar energy densities of rotating ($\bar{r}_{B \textrm{ratio}} = 0.9$) DANS models with the same central energy densities.
\label{fig:density}}
\end{figure}

 In this paper, our goal is to model the final stationary state of a NS with a dark halo that has been spun up through a baryon accretion process which only adds angular momentum to the baryonic fluid, leaving the DM with zero angular momentum, as described in Section~\ref{sec:modelaxi}. However, it is not possible to associate the final stationary state of an axisymmetric DANS with its initial spherically symmetric state without the details of the accretion process. In particular, an accretion model that defines the specific angular momentum of the accreted baryons is required. 
 If the initial stationary spherically symmetric DANS has central energy densities $\epsilon_{B \left( D \right)ci}$, the final stationary axisymmetric DANS will have central energy densities $\epsilon_{B \left( D \right)cf}$ that depend on the number of baryons and amount of angular momentum added to the DANS in the accretion process. In general, the addition of baryons will increase $\epsilon_{Bcf}$ and the addition of angular momentum will decrease $\epsilon_{Bcf}$. 

Our aim is to construct the final axisymmetric rotating DANS after such a baryonic accretion process takes place, so we will not attempt to model the details of the accretion process. In our rotating models, we will choose values for the two central densities that are the same as the central densities of a non-rotating DANS with the same set of EOS such that,
\begin{equation}
\epsilon_{B \left( D \right)cf} = \epsilon_{B \left( D \right)ci} = \epsilon_{B \left( D \right)c}.
\label{eq:epsilonXc}
\end{equation}
In particular, as was done in Section~\ref{sec:modelcompare}, we choose central energy densities such that for the non-rotating DANS, $f_{\chi, \textrm{TOV}} = 0.05$, and compare spherically symmetric and axisymmetric stationary configurations.

\begin{table}[ht!]
\centering
\resizebox{\textwidth}{!}{
\begin{tabular}{cccccccccccccc}
\hline
    colour &
    $m_{\chi}$ &
    $y$ &
    $\epsilon_{Bc}$ &
    $\epsilon_{Dc}$ &
    $\bar{r}_{B \textrm{ratio}}$ &
    $\Omega_{Be}/ \left( 2 \pi \right)$ &
    $\Omega_{De}/ \left( 2 \pi \right)$ &
    $M_{T, \textrm{g}}$ &
    $M_{T, \textrm{g}} \left( R_{B} \right)$ &
    $R_{Be} $ &
    $R_{Bp} $ &
    $R_{De} $ &
    $R_{Dp} $
    \\
    &
    $\left[ \textrm{GeV} \right]$ &
    &
    $\left[ \textrm{MeV/fm\textsuperscript{3}} \right]$ &
    $\left[ \textrm{MeV/fm\textsuperscript{3}} \right]$ &
    &
    $\left[ \textrm{Hz} \right]$ &
    $\left[ \textrm{Hz} \right]$ &
    $\left[ M_{\odot} \right]$ &
    $\left[ M_{\odot} \right]$ &
    $\left[ \textrm{km} \right]$ &
    $\left[ \textrm{km} \right]$ &
    $\left[ \textrm{km} \right]$ &
    $\left[ \textrm{km} \right]$\\
\hline
    black & $0$ & $0$ & $289$ & $0$ & $1$ & $0$ & $0$ & $1.40$ & $1.3984$ & $13.76$ & $13.76$ & $0$ & $0$ \\
    black & $0$ & $0$ & $289$ & $0$ & $0.9$ & $489$ & N/A & $1.48$ & $1.4776$ & $14.48$ & $13.19$ & $0$ & $0$ \\
    yellow & $0.3$ & $0$ & $293$ & $20.79$ & $1$ & $0$ & $0$ & $1.40$ & $1.3936$ & $13.47$ & $13.47$ & $19.38$ & $19.38$ \\
    yellow & $0.3$ & $0$ & $293$ & $20.79$ & $0.9$ & $505$ & $24.4$ & $1.48$ & $1.4718$ & $14.19$ & $12.92$ & $18.40$ & $18.07$ \\
    red & $1$ & $100$ & $284$ & $2.658$ & $1$ & $0$ & $0$ & $1.40$ & $1.3416$ & $13.66$ & $13.66$ & $48.53$ & $48.53$ \\
    red & $1$ & $100$ & $284$ & $2.658$ & $0.9$ & $484$ & $2.47$ & $1.46$ & $1.4197$ & $14.38$ & $13.09$ & $38.75$ & $38.65$ \\
    blue & $1$ & $1000$ & $282$ & $0.03067$ & $1$ & $0$ & $0$ & $1.40$ & $1.3299$ & $13.71$ & $13.71$ & $530.12$ & $530.12$ \\
    blue & $1$ & $1000$ & $282$ & $0.03067$ & $0.9$ & $480$ & $0.0956$ & $1.41$ & $1.4077$ & $14.44$ & $13.14$ & $114.02$ & $114.02$ \\
\hline
\end{tabular}}
\caption{Properties of DANS with central energy densities such that $f_{\chi, \textrm{TOV}} = 0.05$, and a reference pure baryonic NS constructed for comparison. Colour names refer to the colours used for the different DM-EOSs in the figures.\label{tab:densityDANS}}
\end{table}

Figure~\ref{fig:density} shows energy density distributions of DANS as a function of $r$ with $\bar{r}_{B \textrm{ratio}} = 1$ and $0.9$. Various properties of these DANS are listed in Table \ref{tab:densityDANS}. The radial coordinates where the BM (dash-dot) and DM (solid) energy densities go to zero represent $R_{B}$ and $R_{D}$, respectively. In Figure~\ref{fig:density} (left), the baryonic energy densities cannot be visually differentiated due to the log scale of the radial axis. In Figure~\ref{fig:density} (right), we zoom in near the baryonic radius to compare the sizes of $R_{B}$ for the different DANS. A reference pure baryonic NS (black dash-dot) is also shown, to provide a comparison with each of the DANS models.

In Figure~\ref{fig:density} (top) we show spherically symmetric DANS ($\bar{r}_{B \textrm{ratio}} = 1$) with $M_{T} = 1.4 M_{\odot}$ and $f_{\chi, \textrm{TOV}} = 0.05$. These DANS models were previously computed by ref.~\cite{Shawqi2024}. In Figure~\ref{fig:density} (top left), for $m_{\chi} = 0.3$ GeV, $y = 0$, the DANS has a compact halo of $R_{D} = 19.38$ km. The intermediate halo produced with $m_{\chi} = 1$ GeV, $y = 100$, has a dark radius of 
$R_{D} = 48.53$ km. A large diffuse halo of $R_{D} = 530.12$ km is produced for $m_{\chi} = 1$ GeV, $y = 1000$. In Figure~\ref{fig:density} (top right), due to the additional gravitational attraction of the DM within $R_{B}$, the baryonic radius is pulled inwards for the three DANS models compared to the reference pure baryonic NS (black). The smaller the self-interaction strength $y$ is, the larger the amount of DM within $R_{B}$, and thus the more $R_{B}$ is pulled inwards.

For $\bar{r}_{B\textrm{ratio}} = 0.9$, the halo sizes decrease for all the DANS models. Figure~\ref{fig:density} (centre) shows the equatorial $\left( \theta = \pi/2 \right)$ energy density distributions for rotating DANS with the same central energy densities as Figure~\ref{fig:density} (top) and $\bar{r}_{B\textrm{ratio}} = 0.9$. Figure~\ref{fig:density} (centre right) shows $R_{Be}$ increasing for all the stars due to rotation of the baryonic fluid. The increase in $R_{Be}$ compared to spherical symmetry is approximately $5\%$ for all the stars. The increased energy within $R_{B}$ due to rotation causes the dark matter outside $R_{B}$ to be pulled inwards, reducing the sizes of the halos, as can be seen in Figure~\ref{fig:density} (centre left). Compared to the spherically symmetric halos, a smaller dark halo reduces the dark cloud mass $M_{c,\textrm{g}}$ that can be supported. This effect is the most pronounced for the DANS with the largest halo, due to how diffusely the dark matter is distributed in this star, with $R_{De}$ decreasing by $\sim 80\%$, and $M_{c,\textrm{g}}$ decreasing by $\sim 97\%$. For the intermediate halo, the decreases are $\sim 20\%$ for $R_{De}$, and $\sim 39\%$ for $M_{c,\textrm{g}}$. The effect is the lowest for the compact halo with $R_{De}$ decreasing by $\sim 5\%$, and $M_{c,\textrm{g}}$ decreasing by $\sim 37\%$.

\begin{figure}[ht!]
\includegraphics[width=\textwidth]{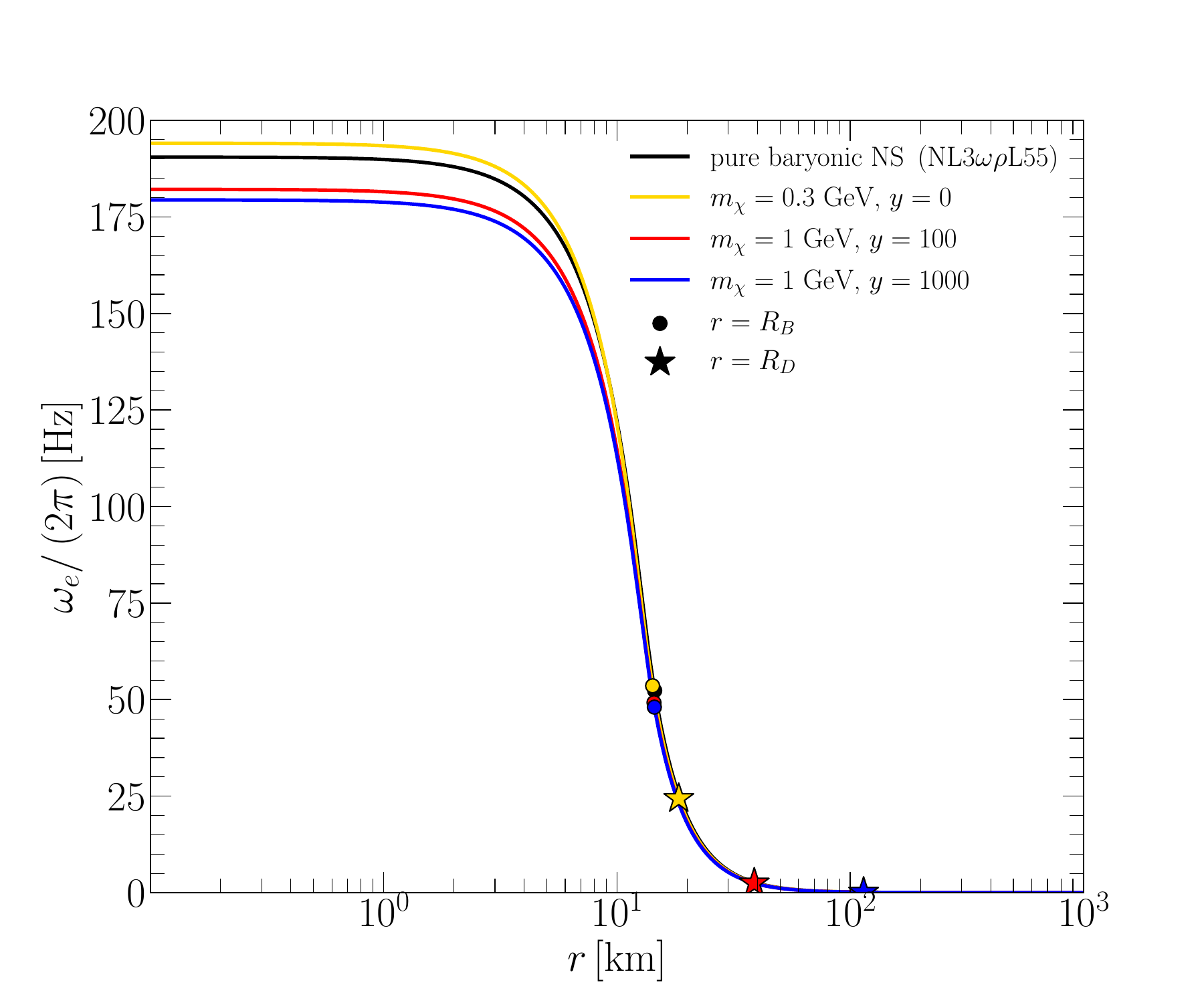}
\caption{Equatorial frame-dragging frequency as a function of radial distance from the centre for the rotating DANS in Table~\ref{tab:densityDANS} with $\bar{r}_{B \textrm{ratio}} = 0.9$. Coloured circles and stars represent the baryonic and dark surfaces of the stars, respectively.
\label{fig:omega}}
\end{figure}

Figure~\ref{fig:density} (bottom) shows the polar $\left( \theta = 0 \right)$ energy density distributions for rotating DANS with the same central energy densities as Figure~\ref{fig:density} (top) and $\bar{r}_{B\textrm{ratio}} = 0.9$. Figure~\ref{fig:density} (bottom right) shows $R_{Bp}$ decreasing for all the stars due to the oblate shape of the star caused by the rotation of the baryonic fluid. The decrease in $R_{Bp}$ compared to spherical symmetry is approximately $4\%$ for all the stars. $R_{Dp}$ also shrinks in size due to spinning of the DANS. The oblate shape is most pronounced for the small halo due to its proximity to $R_{B}$, where the effect of frame-dragging is the highest. For bigger halos, as the dark radius gets further away from $R_{B}$, the effect of frame-dragging decreases, and the dark halos become closer to spherically symmetric, as shown in Figure~\ref{fig:density} (bottom left).


In Figure~\ref{fig:omega}, we plot the equatorial spacetime frame-dragging frequency $\left[ \omega_{e}/ \left( 2 \pi \right) \right]$ as a function of radial distance from the centre, for the rotating DANS listed in Table \ref{tab:densityDANS}. The coloured circles and stars represent the baryonic and dark surfaces of the DANS, respectively. The DANS rotating with $\bar{r}_{B\textrm{ratio}} = 0.9$ leads to equatorial frame-dragging frequencies at the baryonic surfaces of $\omega_{Be}/ \left( 2 \pi \right) \sim 48 - 54$ Hz, with the compact DANS having the highest frame-dragging frequency, followed by the pure baryonic NS, and the DANS with the diffuse halo having the lowest equatorial frame-dragging frequency at the baryonic surface. Since we assume that the dark fluid has zero angular momentum, its spin frequency is equal to the spacetime frame-dragging frequency (Equation~\eqref{eq:darkvel}). With $\omega_{e}$ rapidly falling off with radial distance from the centre, the frame-dragging frequency at the dark surface $\left[ \omega_{De}/ \left( 2 \pi \right) \right]$ is the largest $\left( 24.4 \textrm{ Hz} \right)$ for the smallest halo, and lowest $\left( 0.0956 \textrm{ Hz} \right)$ for the largest halo, with the intermediate halo having $ \omega_{De}/ \left( 2 \pi \right) \sim 2.47 \textrm{ Hz}$. 


\section{Mass-Radius Relations} \label{sec:MR}

\begin{figure}[ht!]
\includegraphics[width=\textwidth]{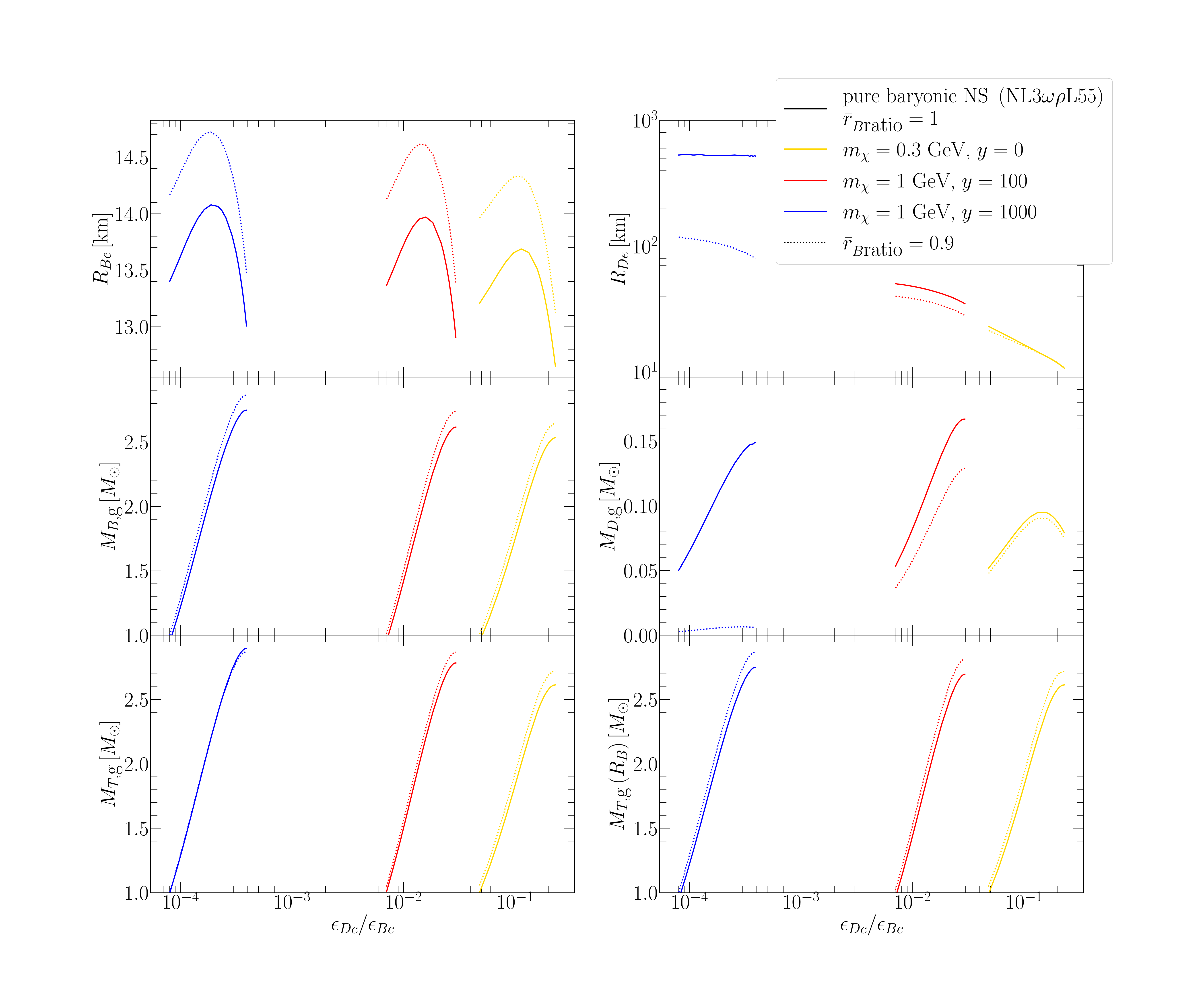}
\caption{Various properties [$R_{Be}$ (top left), $R_{De}$ (top right), $M_{B, \textrm{g}}$ (centre left), $M_{D, \textrm{g}}$ (centre right), $M_{T, \textrm{g}}$ (bottom left), and $M_{T, \textrm{g}} \left( R_{B} \right)$ (bottom right)] for sequences of DANS as functions of $\epsilon_{Dc}/\epsilon_{Bc}$. Solid curves correspond to non-rotating DANS, and dotted curves are rotating DANS with the same central densities as the non-rotating models. All non-rotating models have $f_{\chi, \textrm{TOV}} = 0.05$. 
\label{fig:epsilonMR}}
\end{figure} 


Figure~\ref{fig:epsilonMR} shows various properties for sequences of DANS models, as functions of the ratio of central dark energy density to the central baryonic energy density, $\epsilon_{Dc}/\epsilon_{Bc}$. The solid curves represent stars with $\bar{r}_{B \textrm{ratio}} = 1$, and the dotted curves represent $\bar{r}_{B \textrm{ratio}} = 0.9$. As previously done throughout the paper, the non-rotating stars have $f_{\chi, \textrm{TOV}} = 0.05$ and the rotating stars have the same central energy densities as the non-rotating stars. Figure~\ref{fig:epsilonMR} (top left) plots $R_{Be}$ for the sequences of DANS models. With added rotation, the shapes of the stars become oblate, leading to $R_{Be}$ increasing in size for all the DANS models. Since these stars are rotating rapidly with spin frequencies near 500 Hz, the increase in the equatorial baryonic radius is relatively large, about 0.5 km, similar to the increase in radius for a purely baryonic NS spinning at the same rate.

Figure~\ref{fig:epsilonMR} (top right) shows $R_{De}$ for the same sequences of DANS models. Similar to the cases explained in Section~\ref{sec:density}, $R_{De}$ decreases for all the models for $\bar{r}_{B \textrm{ratio}} = 0.9$. Rotation of the baryonic fluid causes additional gravitational attraction of the DM situated outside $R_{B}$. Since we do not add angular momentum to the DM, the dark fluid is pulled inwards, reducing the size of $R_{D}$. Since the dark fluid is very thinly distributed for the diffuse halos, they experience the largest reductions in size.

As explained in detail in Section~\ref{sec:modelmass}, $M_{B \left( D \right), \textrm{g}}$ should not be interpreted as the actual gravitational mass of each of the fluids separately. However, to understand how each of the terms contribute to the total gravitational mass, $M_{T, \textrm{g}}$, we plot $M_{B, \textrm{g}}$ and $M_{D, \textrm{g}}$ in the centre left and right panels of Figure~\ref{fig:epsilonMR}. $M_{T, \textrm{g}}$ is plotted in the bottom left panel. As can be seen for all cases, the baryonic fluid rotating with the dark fluid remaining at zero angular momentum increases $M_{B, \textrm{g}}$ and decreases $M_{D, \textrm{g}}$. Whether $M_{T, \textrm{g}}$ increases or decreases due to the rotation prescription depends on the relative sizes of the changes in $M_{B, \textrm{g}}$ and $M_{D, \textrm{g}}$. For most cases, the decrease in $M_{D, \textrm{g}}$ is smaller in size than the increase in $M_{B, \textrm{g}}$, and so $M_{T, \textrm{g}}$ increases. However, for the diffuse halos, the decrease in $M_{D, \textrm{g}}$ is the most pronounced and in some cases exceeds the size of the increase in $M_{B, \textrm{g}}$. As a result, $M_{T, \textrm{g}}$ for these diffuse halos is lower for $\bar{r}_{B \textrm{ratio}} = 0.9$ compared to $\bar{r}_{B \textrm{ratio}} = 1$.

Figure~\ref{fig:epsilonMR} (bottom right) plots $M_{T, \textrm{g}} \left( R_{B} \right)$ for the DANS. Since almost all the DM mass resides in the DM clouds for the non-rotating diffuse halos \citep{Shawqi2024}, $M_{T, \textrm{g}} \left( R_{B} \right)$ for the diffuse halos closely resembles their $M_{B, \textrm{g}}$ curves in both the non-rotating and rotating cases. However, significant amounts of DM mass resides within the baryonic surfaces of the non-rotating intermediate and compact DANS \citep{Shawqi2024}, and thus their $M_{T, \textrm{g}} \left( R_{B} \right)$ is larger than their corresponding $M_{B, \textrm{g}}$ curves.

\begin{figure}[ht!]
\includegraphics[width=\textwidth]{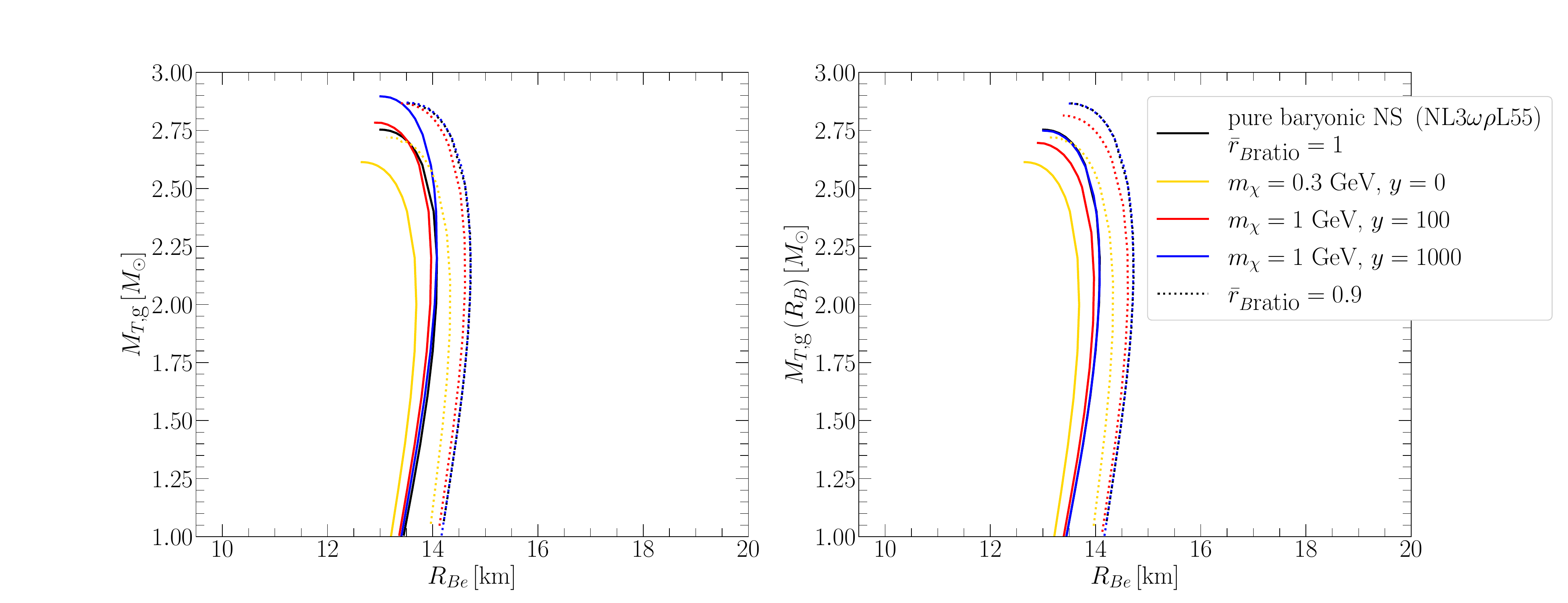}
\caption{Mass versus equatorial baryonic radius curves for the same sequences of stars shown in Figure~\ref{fig:epsilonMR}.
Left: Total gravitational mass as a function of $R_{Be}$. Right: Total gravitational mass enclosed within $R_B$, $M_{T} \left( R_{B} \right)$, for the same DANS models.}
\label{fig:MR}
\end{figure}


Ref.~\cite{Shawqi2024} showed that $M_{T}$, $M_{T} \left( R_{B} \right)$, and $R_{B}$ are the relevant mass-radius variables for measuring masses and radii of NSs with DM halos using electromagnetic observations. For mass measurements using binary orbital dynamics, the relevant mass measured is $M_{T}$. If pulsed X-ray emissions from the baryonic surface are analysed, $M_{T} \left( R_{B} \right)$ and $R_{B}$ are the quantities measured. So in Figure~\ref{fig:MR}, we plot $M_{T, \textrm{g}}$ (left) and $M_{T, \textrm{g}} \left( R_{B} \right)$ (right), as functions of $R_{Be}$, for the same sequences of DANS as in Figure~\ref{fig:epsilonMR}.

As also explained by ref.~\cite{Shawqi2024}, $M_{T} \left( R_{B} \right)$ versus $R_{B}$ curves for stars with DM halos should be directly compared with mass and radius measurements only when the measurement purely involves gravitational self-lensing analysis without use of any prior information about $M_{T}$ measurements coming from radio observations and binary dynamics modeling. To compare with current NS mass and radius measurements, only DM halos that introduce sufficiently low deviations in observed flux can be analysed. Spin frequencies of pulsars are known with very high precision. If mass-radius curves of rotating stars with constant $\bar{r}_{B \textrm{ratio}}$ are plotted (as we do here), the stars on the curves have varying spin frequencies. As a result, it is not appropriate to compare such curves of constant $\bar{r}_{B \textrm{ratio}}$ with the mass and radius measurement of any particular star with a precisely known spin frequency. Thus, we do not show any mass-radius measurements of NSs in Figure~\ref{fig:MR}.

At present, all EOS inference codes (e.g. \citep{2019MNRAS.485.5363Greif}) solve the TOV equations and do not take into account rotation. This is a valid approach since the pulsars presently observed by NICER rotate slowly, with spin rates near 200 Hz. Universal relations for the change in mass and radius for pure neutron stars \citep{Konstantinou2022} could be employed to include rotation if more rapidly rotating NSs are observed in the future. The universal relations for purely baryonic NS can be extended to core DANS \citep{Konstantinou2024}, but whether they hold for halo DANS is still an open question that could be examined in future work.

\section{Deviation in Metric Compared to No-Halo Schwarzschild Counterpart} \label{sec:metric}

\begin{figure}[ht!]
\centering
\includegraphics[width=\textwidth]{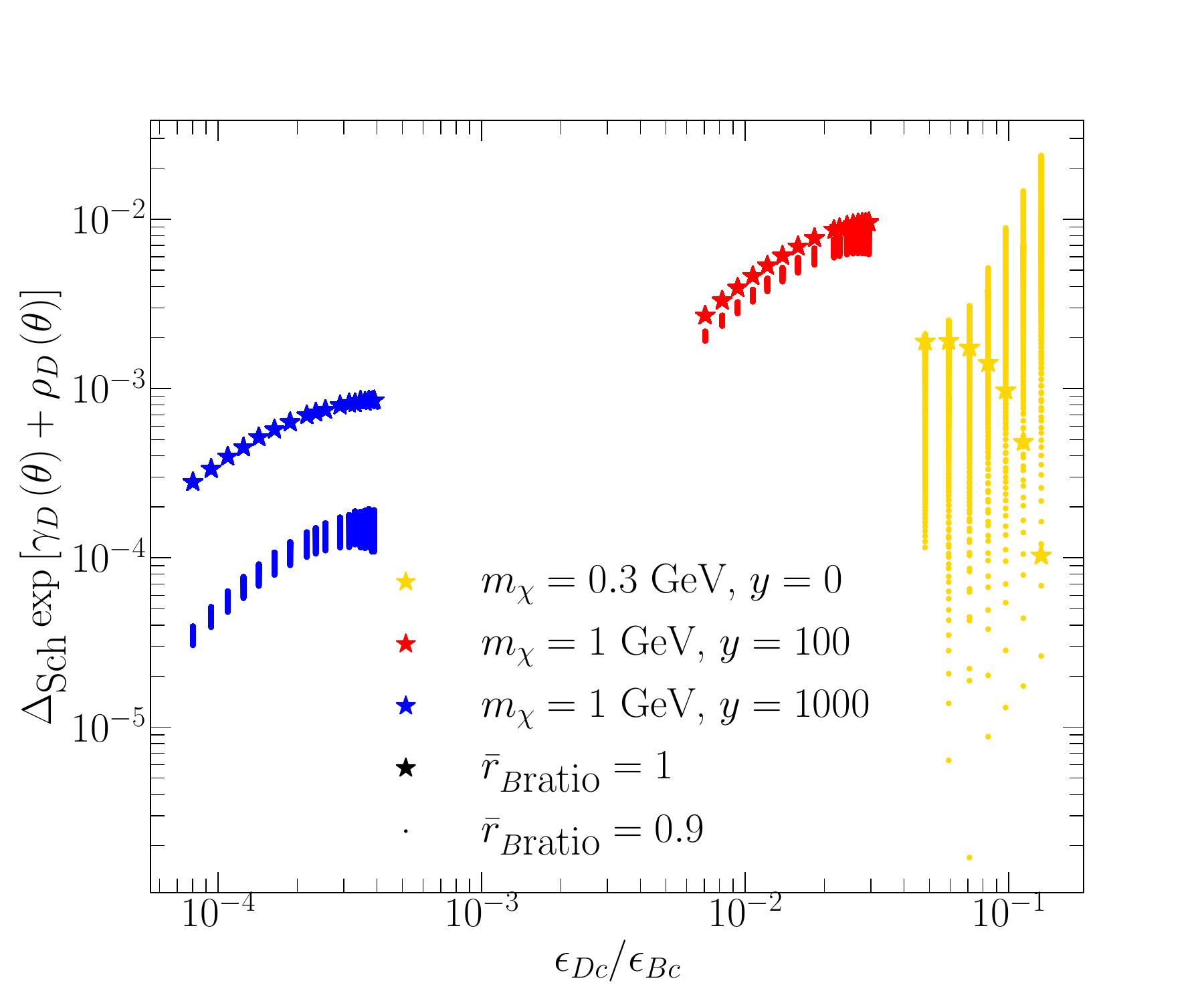}
\caption{Equation~\eqref{eq:Deltametricaxi} as a function of ratios of central DM energy densities to central BM energy densities for the same sequence of DANS shown in Figures \ref{fig:epsilonMR} and \ref{fig:MR}. Star symbols represent non-rotating DANS. Dots correspond to varying $\theta$ on the DM surfaces of rotating DANS.
\label{fig:metricdiff}}
\end{figure}

If a DM cloud exists outside the baryonic surface of a NS, the paths of X-rays emitted from the NS baryonic surface will deviate compared to those emitted from a NS with no DM cloud. This will alter the X-ray flux observed by telescopes, such as Chandra and NICER, due to the additional gravitational lensing caused by the DM cloud. Present-day mass and radius inference using observed X-ray flux assumes a Schwarzschild vacuum spacetime outside the baryonic surface of NSs. Thus, if a DANS has a dark halo that introduces significant deviations in the observed flux (compared to a NS with no DM), conclusions drawn from mass and radius inference will be incorrect. 

However, in some cases, the dark matter in the cloud is very diffuse. In these diffuse DM cloud models,
the DM in the dark cloud causes negligible deviations in the path of X-rays, and the spacetime outside the baryonic surface can be well approximated by the Schwarzschild metric \citep{Miao2022, Shawqi2024}. Only in these cases can NICER mass and radius inferences be used to constrain DANS with DM halos.

Ref.~\cite{Miao2022} showed how to identify this subset of halos in spherical symmetry, through the introduction of a \textit{no-halo Schwarzschild counterpart} (NHSC) to a halo DANS. The NHSC is a reference star with no DM cloud, with a total TOV mass that is the same as $m_{T} \left( R_{B} \right)$ of its DANS counterpart. However, it should be noted that the NHSC is not a solution of the TOV equations with the same EOS as the baryonic matter in the DANS. Ref~\cite{Miao2022} found that when comparing a NHSC with a halo DANS, there is a power-law relation between the change in the time-time component of the spacetime metric evaluated at $r = R_{D}$ and the magnitude of the maximum relative change in flux. Thus, given any threshold maximum allowed change in flux, the power law dictates the maximum allowed change in metric at the outer edge of the halo. The threshold maximum flux change depends on the statistics of the observation \citep{Shawqi2024}.



The X-ray pulsars observed by NICER are rotating. The most important features of rotation are introduced into the analysis of NICER data through the use of the simple Oblate Schwarzschild (OS) approximation \citep{Morsink2007}. In the OS approximation, Doppler terms are introduced in the same way as in the Schwarzschild plus Doppler approximation \citep{Miller1998,Poutanen2003}, and the oblate shape is introduced as a parametrized surface embedded in the Schwarzschild metric. The parametrized surface depends in a quasi-universal manner on the star's mass, equatorial radius, and spin frequency. 

In order to extend the OS approximation to rotating DANS, we have identified two necessary conditions. First, an extension of the quasi-universal shape parametrization to DANS is required. If this shape function is the same as for purely baryonic neutron stars, then the adaptation of NICER results may be possible. The second condition is that the presence of a dark matter halo should not alter the deflection of photons and the observed X-ray pulse profiles by more than some threshold amount. The case of DANS with DM cores was investigated by ref.~\cite{Konstantinou2024}, who found that the quasi-universal shape functions are unchanged by a dark matter core. This means that NICER results can be adapted to the restricted family of DANS with DM cores \citep{Rutherford2023,Rutherford2024}.

The effect of DM halos on the oblate shape and the gravitational lensing of DANS is more complex than the case of DM cores, and a full examination of these changes is beyond the scope of this paper. However, we can make some simple hypotheses on how the OS approximation could be adapted to the study of DANS with DM halos. It is natural to make the ansatz that a generalized NHSC could be constructed for a rotating DANS, making use of the global definition of mass.  
If an axisymmetric DANS with a dark halo has total mass enclosed within its baryonic radius, $M_{T, \textrm{g}} \left( R_{B} \right)$, then our ansatz is that the NHSC has total mass $M_{T, \textrm{g}} = M_{T, \textrm{g}} \left( R_{B} \right)$, and no DM cloud. In spherical symmetry, the NHSC star has a radius that is the same size as the baryonic radius of its DANS counterpart. For a rotating DANS, the NHSC star has an oblate surface that is the same shape and size as the baryonic surface of its DANS counterpart. The spacetime outside the baryonic surface of the NHSC is a Schwarzschild vacuum in both spherical symmetry and axisymmetry.


Calculating observed flux from DANS through raytracing in axisymmetric spacetimes is out of the scope of this paper. Instead, we compare the deviations in metric of the DANS at $r = R_{D}$, compared to their respective no-halo Schwarzschild counterparts. In spherical symmetry, the absolute difference in the time-time component of the metric at $r = R_{D}$, between a DANS with a dark halo and its no-halo Schwarzschild counterpart is
\begin{equation}
\left| 1 - \frac{2 M_{T, \textrm{TOV}}}{R_{D}} - \left[ 1 - \frac{2 m_{T} \left( R_{B} \right)}{R_{D}} \right] \right| = \frac{2 M_{c, \textrm{TOV}}}{R_{D}}.
\label{eq:Deltametricspher}
\end{equation}
In axisymmetry, due to the oblate shape of the DANS, $R_{D}$, as well as the metric evaluated at $r = R_{D}$, are functions of $\theta$. Let $\gamma_{D} \left( \theta \right)$ and $\rho_{D} \left( \theta \right)$ be the metric potentials evaluated at $R_{D} \left( \theta \right)$. Then define the function
\begin{equation}
\Delta_{\textrm{Sch}} \exp{\left[ \gamma_{D} \left( \theta \right) + \rho_{D} \left( \theta \right) \right]} = \left| \exp{\left[ \gamma_{D} \left( \theta \right) + \rho_{D} \left( \theta \right) \right]} - \left[ 1 - \frac{2 M_{T, \textrm{g}} \left( R_{B} \right)}{R_{D} \left( \theta \right)} \right] \right|.
\label{eq:Deltametricaxi}
\end{equation}
In Figure~\ref{fig:metricdiff} we compute $\Delta_{\textrm{Sch}} \exp{\left[ \gamma_{D} \left( \theta \right) + \rho_{D} \left( \theta \right) \right]}$ as a function of $\epsilon_{Dc}/\epsilon_{Bc}$. The spherically symmetric cases are represented by star symbols, and the rotating stars are represented by dots. For a particular choice of $m_{\chi}$ and $y$, the deviations increase with increasing $\epsilon_{Dc}/\epsilon_{Bc}$. In Figure~\ref{fig:epsilonMR} (bottom left), it was shown that $M_{T, \textrm{g}}$ increases with $\epsilon_{Dc}/\epsilon_{Bc}$. Thus, the maximum deviations in the metric compared to the no-halo Schwarzschild counterparts increase with increasing $M_{T, \textrm{g}}$. For the large halos, the deviations in axisymmetry are smaller than those in spherical symmetry. This is because the mass of the dark cloud is much smaller in axisymmetry, as described in Section~\ref{sec:density}. While the maximum deviations for the compact halos increase with $M_{T, \textrm{g}}$ in axisymmetry, the deviations in spherical symmetry decrease with increasing $M_{T}$. This is because the compact halos have smaller and smaller masses in their clouds with increasing $M_{T, \textrm{g}}$.

It should be emphasized that we have not proved that deviation in observed flux from DANS with dark halos, compared to their no-halo Schwarzschild counterparts, is related to $\Delta_{\textrm{Sch}} \exp{\left[ \gamma_{D} \left( \theta \right) + \rho_{D} \left( \theta \right) \right]}$. Since a similar empirical relation exists in spherical symmetry \citep{Miao2022}, we are hypothesizing that this quantity could be an indicator of whether observed flux from a dark halo can be well approximated by an oblate no-halo Schwarzschild counterpart embedded in a Schwarzschild spacetime. Evaluating whether this hypothesis is true requires calculation of observed flux through raytracing, and is left for future work.

\section{Discussion and Conclusions} \label{sec:conclusions}

Rotation causes a NS to deform into an oblate shape. If DM exists in NSs, both the baryonic and dark components will deform into oblate shapes due to rotation. In this paper, we compute the structures of rapidly rotating DANS
using a two-fluid modification of the {\tt{RNS}} code, focusing on dark halo configurations.

Our motivation is to describe recycled pulsars spun up from a zero angular momentum state by baryonic matter accreted from a companion star. The end state of such a process will be rigid rotation for the BM, just as occurs for purely baryonic NSs. Meanwhile, the dark fluid does not gain angular momentum, with the result that the dark fluid gains a differential angular velocity equal to the frame-dragging of spacetime caused by the baryonic fluid's rotation. Our tacit assumption here is that it is unlikely that there is any physical mechanism that would enforce the rigid rotation of the DM. Microscopic computations of the viscosity of self-interacting DM would be useful to test this assumption, but at present, we are unaware of any such result.

Due to the importance of mass in astronomy, we also consider the local and global general relativistic definitions of mass in spherical symmetry. For non-rotating DANS, these two definitions agree when the total mass is computed. It is common to split the DANS mass into baryonic and dark components using the local (TOV) definition of mass for non-rotating DANS. However, only the global definition of mass exists for rotating DANS, so it is useful to explore the differences in these definitions. We emphasize that neither local nor global definitions of $M_{B}$ and $M_{D}$ are physically measurable invariant quantities, and neither should be interpreted as the gravitational mass of each of the respective fluids, even for non-rotating stars. We find that the differences in the local and global definitions of the BM and DM masses are small, but numerically resolvable for the DANS computed in this paper. This means that the DM mass fraction, $f_\chi$ differs depending on the choice of definition for spherical stars. In our calculations with $f_{\chi, \textrm{TOV}} = 0.05$ we find that the DM mass fraction could change by at most 10\% if the global definition of DM mass were used instead. 

In our computations of rotating DANS, we restrict our attention to one baryonic EOS and three types of fermionic DM EOS that lead to three types of halos: compact, intermediate, and diffuse.
When we compare two DANS with the same EOSs and central densities, we find that rotation causes the baryonic equatorial radius to increase and the polar radius to decrease, similar to the situation for pure baryonic NSs. 
The increased energy within $R_{B}$ due to rotation pulls the dark radius inwards, shrinking the size of $R_{D}$ compared to the spherically symmetric cases for all three types of halos. This effect is strongest for the largest diffuse halos. The dark halos are also less oblate than the baryonic surfaces.

We compute the baryonic $M_{B, \textrm{g}}$  and DM components of the DANS mass $M_{D, \textrm{g}}$ using the global mass definition to understand the variation of total mass of the DANS. For all halos, $M_{B, \textrm{g}}$ increases and $M_{D, \textrm{g}}$ decreases due to rotation, compared to the spherically symmetric cases with the same central densities.
For most halos, the size of the increase in $M_{B, \textrm{g}}$ is greater than the size of the decrease in $M_{D, \textrm{g}}$, leading to an overall increase in $M_{T, \textrm{g}}$ compared to the spherically symmetric case. But for some diffuse halos, the size of the decrease in $M_{D, \textrm{g}}$ is larger, leading to an overall decrease in $M_{T, \textrm{g}}$ due to rotation.

Before an X-ray telescope, such as NICER, can detect the effects of a DM halo in a pulsar, some issues still need to be resolved. The analysis of X-ray data uses quasi-universal empirical relations for the oblate shape of the baryonic surface. It will be necessary to investigate the changes in the shape caused by a DM halo. It is also important to compute the extra deflection of photons caused by a DM halo that exists around a rotating DANS. 
The resolution of both questions could be answered in the future using the code described in this paper.

The C program used to generate results in this paper can be found at \url{https://github.com/rns-alberta/TwoRNS}.

\acknowledgments

This research was supported in part by AGES scholarship awarded to S.S., and NSERC Discovery Grant RGPIN-2019-06077.




\bibliographystyle{JHEP}
\bibliography{biblio.bib}
\end{document}